\documentclass[preprint]{aastex}

\begin{document}

\title{The Tolman Surface Brightness Test for the Reality of the
Expansion. I. Calibration of the Necessary Local Parameters}

\author{Allan Sandage} 
\affil{Observatories of the Carnegie Institution of Washington,\\ 813
Santa Barbara Street, Pasadena, California 91101}
 
\author{Lori M. Lubin\altaffilmark{1,2}} \affil{Department of Astronomy,
California Institute of Technology,\\ Mailstop 105-24, Pasadena,
California 91125}

\altaffiltext{1}{Hubble Fellow} \altaffiltext{2}{Current address :
Department of Physics and Astronomy, Johns Hopkins University,
Baltimore, MD 21218}

\begin{abstract}

The extensive CCD photometry by Postman \& Lauer (1995, ApJ, 440, 28)
in the Cape/Cousins $R$ photometric band for first ranked cluster
elliptical and S0 galaxies in 118 low redshift ($\langle z \rangle =
0.037$) clusters is analyzed for the correlations between average
surface brightness, linear radius, and absolute magnitude. The purpose
is to calibrate the correlations between these three parameters in the
limit of zero redshift. The Postman-Lauer cluster galaxies at low
redshift approximate this limit. We apply small corrections for the
finite mean redshift of the sample in order to define the
zero-redshift correlations.  These local correlations provide the
comparisons to be made in Paper IV with the sample of early-type
galaxies at high redshift in search of the Tolman surface brightness
signal of $(1 + z)^4$ if the expansion is real.

Surface brightness averages are calculated at various metric radii in
each galaxy in the sample. The definition of such radii by Petrosian
(1976, ApJ, 209, L1) uses ratios of observed surface photometric data.
Petrosian radii have important properties for the Tolman test which
are reviewed in this paper. The observed surface brightnesses are
listed for 118 first ranked cluster galaxies at Petrosian $\eta$ radii
of 1.0, 1.3, 1.5, 1.7, 2.0, and 2.5 mag.  The three local diagnostic
correlation diagrams are defined and discussed. We review the Tolman
test and show that, although recipes from the standard cosmological
model that already have the Tolman signal incorporated are required to
calculate linear radii and absolute magnitudes from the observed data,
the test is nevertheless free from the hermeneutical circularity
dilemma occasionally claimed in the literature. The reasons are the
observed mean surface brightness (1) is independent of any assumptions
of cosmological model, (2) does not depend on the existence of a
Tolman signal because it is calculated directly from the data using
only angular radii and apparent magnitudes, and (3) can be used to
search for the Tolman signal because it carries the bulk of that
signal.

\end{abstract}


\section{Introduction}

Tolman (1930, 1934) derived the remarkable result that in an expanding
universe with any arbitrary geometry, the surface brightness of a set
of "standard" (identical) objects will decrease by $(1 + z)^4$. One
factor of $(1 + z)$ comes from the decrease in the energy of each
photon due to the redshift. The second factor comes from the decrease
in the number-flux per unit time. Two additional factors of $(1 + z)$
come from the apparent increase of area due to aberration. The effect
is the same for all intrinsic geometries because the cosmological
geometric effects due to different space curvatures (i.e.\ the
dependence on $q_o$) are identical in the equations for luminosity, $L
= f(q_o, z)$, and intrinsic radius = $g(q_o, z)$.  Hence, the ratio of
$L$ to (radius)$^2$, which is the surface brightness, is independent
of all the cosmological parameters, precisely (Sandage 1961, 1972).

It was realized early that the Tolman prediction constituted a test of
the reality of the expansion because the effect would be different in
an non-expanding universe. There, the redshift is presumed to be due
either to some unknown property of matter or to interactions with
matter in the path length. The surface brightness would then decrease
by only one factor of $(1 + z)$.  Only the energy effect of individual
redshifted photons would be present. The aberration and the decrease
in the flux of energy flow due to the "number effect" would be zero
(Geller \& Peebles 1972; Sandage 1974; Sandage \& Perelmuter 1990a).

Although the test has been known in principle for 70 years, attempts
to implement it have been few because four difficult, practical
problems must first be solved.

\newcounter{discnt}
 
\begin{list}
{\arabic{discnt}.}  {\usecounter{discnt}}

\item How does one use an operationally robust definition of average
surface brightness for any galaxy class, say elliptical and S0
galaxies, when the surface brightness of such galaxies is a strong
function of position in the image, varying by a factor of 1000 between
center and "edge"?

\item How can one define an appropriate radius over which to measure a
mean surface brightness in objects with such strong intensity
gradients? Further, how does one define a {\it metric} size rather
than an {\it isophotal} size that is independent of the cosmological
geometry using only the observed photometric data for the test
galaxies? Use of isophotal radii causes a degeneracy for the test
(Sandage 1961, 1972, 1995).

\item How does one avoid systematic effects at large redshifts due to
the finite resolution of the point-spread functions for the telescopes
used to obtain the data? Angular resolutions of the order of
0\farcs{1} for redshifts larger than $z = 0.5$ are required.

\item How does one account for any evolutionary change in surface
brightness and/or metric size between the high redshift galaxies and
the local galaxies that are used for the comparison due to the
appreciable difference in the look-back times?

\end{list}

Problems 1--3 were addressed in three earlier papers (Sandage \&
Perelmuter 1990a,b, 1991; hereafter SP I, II, and III) where a Tolman
signal was claimed, based on a particular set of ground- based data
(Djorgovski \& Spinrad 1981). However, the results of this test were
marginal because the angular resolution of the ground-based data was
only on the order of 1\farcs{0} and because only photographic data
were used. One of the proposed cosmological projects for a large space
telescope then in the early planning stage (called the LST at the
time) was the Tolman test (Sandage 1974) because the diffraction limit
of the LST would be between 0\farcs{05} and 0\farcs{10}, at least 10
times better than from the ground.  After 25 years, we now make a
first attempt to carry out the test using data from the {\it Hubble
Space Telescope} (HST) for three galaxy clusters at high redshift
studied in a series of papers using the Keck, the Kitt Peak, and the
HST telescopes (Oke, Postman \& Lubin 1998; Postman, Lubin \& Oke
1998, 2001; Lubin et al.\ 1998, 2001). These authors have studied nine
remote clusters, obtaining spectroscopic redshifts for 892 very faint
galaxies with the Keck 10-m telescopes, and $BVRIK$ colors of many
galaxies in the fields with both the Keck and the HST.

We have used the Keck and HST spectroscopic and photometric data for
three of the clusters (Cl 1324+3011, Cl 1604+4304, and Cl 1604+4321).
The redshifts of the clusters are 0.76, 0.90, and 0.92,
respectively. We have determined intensity profiles with radii from
the HST WFPC2 images where there is sufficient angular resolution to
permit reliable Petrosian (1976) angular radii to be measured at
Petrosian $\eta$ values from 1.3 to 2.0 mag. From these data we have
calculated observed surface brightnesses at each of the Petrosian
radii, which, when corrected to rest wavelengths by calculated $K$
terms, permit comparison of surface brightnesses of the high-redshift
cluster galaxies with similar data for local elliptical (E) and S0
galaxies of known absolute magnitudes.

The present paper is the first of four on this Tolman test using HST
data. It is concerned with the calibration of three diagnostic
diagrams that define the relations between surface brightness,
absolute magnitude, and radii for local early-type galaxies. There are
systematic variations between each of these quantities for the
manifold of early-type galaxies. These variations must be calibrated
out to recover a pure Tolman signal from the observations of the
surface brightness of similar high-redshift galaxies as modified by
evolution in the look-back time.

The problem is the same as was addressed in SP II where the
calibrations of the necessary scaling relations were made using
galaxies in the Virgo, Fornax, and Coma clusters. The calibrations
were extended in SP III using additional photographic photometry data
on first, second, and third-ranked galaxies in 56 Abell clusters and
well-known groups. The data used there were from the surface
photometry principally by Oemler (1976), Thuan \& Romanishin (1981),
Malumuth \& Kirshner (1985), and Schombert (1986, 1987).

We have repeated these scaling calibrations in the present paper using
the CCD photoelectric photometry of Postman \& Lauer (1995, hereafter
PL) for the first-ranked galaxies in 118 Abell clusters. The reasons
for repeating the local calibrations are (1) to test the systematic
accuracy of the data in SP I, II, and III using the increased
systematic CCD photoelectric accuracy of the PL data, and (2) to use
the PL data in the $R$ band rather than $B$ and $V$ as in the SP
series. The observations of the high-redshift galaxies with HST were
made in the bandpasses of $F702W$ and $F814W$ which are closer to the
$R$ and $I$ bands than the $V$ band used in SP II and III.

In this first paper of the present series, we set out properties of
the Petrosian $\eta$ metric radii and show again why they are so
important for the test. The calibration of the three correlations of
mean surface brightness, linear radii, and absolute magnitude using
Petrosian metric sizes are the main results of the present paper.  In
Lubin \& Sandage (2001a; hereafter Paper II) we show the sensitivity
of the accuracies of Petrosian radii to the size of the point-spread
functions for both the Keck ground-based and the HST data. We also
address the proper way to reduce profile and surface-brightness data
for highly flattened E galaxies using either circular apertures with
$\sqrt{ab}$ effective radii, where $a$ and $b$ are the semi-major and
semi-minor axes of the best fitting ellipse, or elliptical apertures
with the observed ellipticity.

In Lubin \& Sandage (2001b; hereafter Paper III) we determine the
profile and surface brightness data for the early-type galaxies in the
three high-redshift clusters used here for the test.  Lubin \& Sandage
(2001c; hereafter Paper IV) use the data from Paper III to make the
Tolman test. We discuss there (1) the effects of luminosity evolution
in the look-back time, and (2) what may appear to be a partial
degeneracy of the test where we need knowledge of absolute magnitudes
and linear radii that can only come from the Mattig (1958) equations
at high redshift. These equations already are based on the standard
cosmological model that have the Tolman surface brightness signal
built into them. We show in Paper IV that the test is, however, robust
and that the observed data are consistent with the Tolman prediction
whereas the non-expanding (tired light) models for non-expansion are
not. In \S 5 of the present paper we argue that the test in Paper IV
is not circular because the bulk of the Tolman signal is contained in
the observed surface brightnesses with only a slight dependence on the
linear radii that must be calculated from the assumed cosmology.
 
\section{Petrosian $\eta$ Metric Radii}

\subsection{Definition of $\eta$ and Properties of the Petrosian $\eta$ Ratio}

Definitions and properties of Petrosian metric radii are derived
elsewhere (Djorgovski \& Spinrad 1981, hereafter DS; SP I; Sandage
1995) based either on the intensity profile or the observed growth
curve of magnitude versus aperture. We give here only a summary of the
definitions and some of the remarkable properties of the function.
Defined as twice the slope of the growth curve expressed in magnitudes
as,

\begin{equation}
     \eta ({\rm mag}) = 2.5~{\rm log}~[2d({\rm log}~r)/d\{{\rm log}~L(r)\}]
\end{equation}

\noindent It is proved in DS, SP I, and originally by Petrosian (1976)
that this definition is identical to finding the difference in
magnitude between the mean surface brightness, $\langle SB(r)\rangle$,
averaged over the area interior to a particular radius and the surface
brightness, $SB(r)$, (i.e.\ the profile intensity) at that
radius. Hence, to find the radius where the surface brightness
averaged over the area encompassed by a particular radius is, say, two
magnitudes brighter than the surface brightness at that radius, we
need only calculate the mean surface brightness for various radii and
interpolate to the radius where $\langle SB \rangle$ is two magnitudes
brighter than $SB(r)$.

The calculations can either be made from a known growth curve using
equation (1) if we have aperture photometry with increasing aperture
sizes or by integrating a known intensity profile to calculate the
mean surface brightness at a series of radii. The integral method
using observed profiles was used in SP I. The profile data were
obtained from the literature cited in the last section. The equations
in SP I for $\eta$ are their equation (6) for the definition, equation
(7) for the $\langle SB \rangle$ averaged over $r(\eta)$, equation
(10) for the profile giving the $SB(r)$ at $r(\eta)$, and equation
(13) for $\eta$ expressed in magnitudes.

Because $\eta$ is the {\it ratio} of various surface brightnesses in a
given galaxy, it defines a {\it metric} (rather than an isophotal)
radius that is independent of reddening, absorption, redshift, color
(if there is no color gradient across the image), luminosity evolution
(again if there is no differential evolution across the image), and
the $K$-correction term for the effect of redshift.

\subsection{The Two Ways of Calculating $\eta$}

The method of calculating $\eta$ using only the profile is shown in
Figure 1, taken from Figure 5 of SP I. The profile data for NGC 3379
are from de Vaucouleurs \& Capaccioli (1979). The intensity contained
within each tabulated radius is obtained by numerical integration
using ``circular apertures.'' At each radius the difference in
magnitude between this running integration and the profile value,
$SB(r)$, is calculated. This function is shown in Figure 1 and labeled
as $\eta(r)$. Note the ragged nature of the $\eta$ curve, calculated
there at every listed radius for which the $SB(r)$ profile was given
by de Vaucouleurs \& Capaccili (1979). The growth curve, also
calculated by integrating the profile within circular apertures, is
labeled $m(r)$ in Figure 1.

The calculation of $\eta(r)$ from the slope of the growth curve is
often easier than from the profile because the necessary data are
contained directly from the aperture photometry. Such data were
routinely obtained as the output of the single channel photometers of
the 1950's to 1980's by using various blocking apertures and measuring
the magnitude contained within a particular aperture. The various
photometric programs for the Hubble diagram in the 1970's on cluster
galaxies (e.g.\ Sandage 1972, 1975) were done that way. No knowledge
is needed of the profile, which is, of course, the first derivative of
the growth curve.

Growth curve data are also given by Postman \& Lauer (1995) in their
``aperture'' photometry of first ranked galaxies in the 119 nearest
Abell clusters. Although they used a CCD areal detector, they list the
magnitude inside various circular apertures made by summing the
intensities, pixel by pixel, to given radial distances from the image
center. Because of the straightforward calculation of the slope of the
growth curve directly from the data listed by PL, we have used
equation (1) to calculate $\eta(r)$ values from these data.

Adopting the slope of that curve to be the $\alpha$ parameter of
Postman \& Lauer (1995) which is calculated from the growth curve as

\begin{equation}
             \alpha = 0.4~d[m(r)]/d({\rm log}~r),
\end{equation}
     
\noindent $\eta$ follows from

\begin{equation}
               \eta({\rm mag}) = 2.5~{\rm log}~(\alpha/2),
\end{equation}

\noindent using equation (1). 

\section{The $\eta$ Parameters from the Postman--Lauer Data for
Local First-Ranked Elliptical and S0 Cluster Galaxies}

\subsection{Color and Magnitude Systems}

The Postman \& Lauer (1995) CCD photometric data in the $B$ and $R$
passbands is given in their Table 3. The photometry in $B$ is on the
standard Johnson/Morgan $UBV$ system. The photometry in $R$ is on the
Cape-Cousins $R_C$ system which differs by an appreciable color
equation from the Johnson (1965, 1966 et al.; also Mendoza 1967) $R_J$
system (Sandage 1997, eqs. 1 and 2). That system, $R_J$, is the same
as the original $r(S20)$ system of Sandage \& Smith (1963).  $R_J$ is
also the system used for all the Palomar photometry by one of us for
the bright cluster galaxies, based on the defining list of standards
for that photometry (Sandage 1972, 1973b). The non-linear relation
between $R_J$ and $R_C$ is given elsewhere (Sandage 1997).  

In all the remaining parts of the present series we use the $R_C$
(Cape/Landolt) system as given in the PL photometry based on Landolt
(1983, 1992) standards for the Cape $(RI)_{\rm Cape}$ system.
However, it must be noted in the comparisons of the Postman \& Lauer
(1995) $R_C$ photometry with the $R_{S20}$ photometry of Sandage
(1973b) that the difference in the two $R$ systems at the color of
elliptical galaxies has a zero point offset of 0.26 mag. The $R_C$ is
fainter than $R_J$, and, therefore, the $R$ magnitudes in PL are
fainter by the same amount at all their aperture sizes from the $R$
magnitudes listed by Sandage (1973b) for the galaxies in common.

\subsection{The Linear Radius, Absolute Magnitude, and Surface Brightness at Various $\eta$ Radii for Local First Ranked Cluster Galaxies}

Postman \& Lauer (1995) list their equivalent ``circular'' aperture
photometry for each of their program galaxies using equal intervals of
(log $r''$). They have set their second listed entry for each galaxy
at their standard metric radius of 16 kpc (defined with their adopted
Hubble constant). With equal intervals of log $r''$, the slope of the
growth curve, $\alpha$, is calculated from equation (2) from their
radii and magnitude differences directly. $\eta(r)$ at each of the
listed radii then follows from equation (3).

We calculated the slope of the growth curve at each listing in the
master Postman-Lauer table. This gives the value of $\eta$ at each
listed radius. Interpolation gives the radii for $\eta$ values of 1.0,
1.3, 1.5, 1.7, 2.0 and 2.5 mag. These are the same intervals of $\eta$
used in SP I, II, and III.  The magnitude inside each of the
interpolated $\eta$ values was then read from the plotted growth
curves for each of the 118 PL clusters that have adequate data. The
surface brightness, in mag per arcsec$^2$, is then calculated from,

\begin{equation}
    {\langle SB \rangle}_{\eta} = 2.5~{\rm log}~(\pi r_{\eta}^{2}) + m_{\eta}
\end{equation}

\noindent where $m_{\eta}$ is the apparent magnitude of the light
encompassed within the respective $\eta$ radii, $r_\eta$ (in
arcseconds). Note that this is independent of all cosmology as to
``proper distance.'' It is a {\it directly observed} quantity, no
matter what its interpretation (see \S 3.4).

\subsection{Correction of the Observed Postman--Lauer Data for Galactic Absorption and the $K$ term for the Effect of Redshift} 

The observed apparent magnitudes have been corrected for Galactic
absorption and the $K(z)$ effect of redshift. We first compared the
corrections adopted by Postman \& Lauer (1995) with previous
determinations of the Galactic absorptions, $A(R)$, and the $K(z)$
terms in $R$. The comparison of $K_z (R)$ was with the $K(z)$ values
used in the earlier photometry of clusters (Sandage 1973a, Table
5). The agreement is excellent for both $K_z (R)$ and $K_z (B-R)$ at
the level of 0.01 mag.

The $B-R$ colors listed by Postman-Lauer, corrected by their adopted
$K(B-R)$ values, were then correlated with the cosecant of the
Galactic latitude and with the $E(B-V)$ reddenings adopted by Lauer \&
Postman (1994) from the tables of Burnstein \& Heiles (1984). These
reddenings were also correlated with the cosecant of the Galactic
latitude using the listings of the Galactic coordinates of the
clusters in Table 1 of Lauer \& Postman (1994). These correlations are
highly consistent with the reddening and absorption model of zero
absorption in the pole down to Galactic latitude $45^{\rm o}$, and
then an increase in $A_R$ by 0.08 mag per unit change in cosecant $b$
for csc $>$ 1.5 adopted in Paper VIII of the velocity-distance series
(Sandage 1975). The slope coefficient adopted by Postman \& Lauer
(1995) is 0.09 mag/csc $b$.  In view of the excellent agreement of
these data we have adopted the $K(z)$ and $A_R$ corrections of Postman
\& Lauer (1995).  The adopted corrections for absorption are $A_R =
2.35~E(B-V)$, where we have used the reddenings of Lauer \& Postman
(1994).

\subsection{Listing of the Absolute Magnitudes, Linear Radii, and Observed Surface Brightnesses Corrected for Galactic Absorption and $K(z)$}

The size of the database is very large if we were to include the
complete data on the apparent magnitude and angular radius for each
$\eta$ value for each cluster. We do not list these observed data
here. However, they can be recovered from Table 1 using the listed
absolute magnitudes and linear radii, together with the listed $m - M$
moduli and the factor, $f$, that changes angular radii into linear
radii, as described below. The data in Table 1 were calculated from
the observed apparent magnitudes (corrected for Galactic absorption
and $K$ term) and angular radii (in arcseconds) in the following way.

Each Abell cluster is identified in the extreme left hand column of
Table 1. Under the identification Abell cluster number in that column
is the adopted cluster redshift from Table 1 of Lauer \& Postman
(1994). The adopted distance modulus $m - M$ is given next in that
column, based on the redshift and an arbitrary Hubble constant of 50
km sec$^{-1}$ Mpc$^{-1}$. If we use the naive formulation of the
velocity distance relation (with no regard for the geometry of
different $q_o$ values) where the present distance is simply $cz/H_o$,
then for $H_o = 50$

\begin{equation}
            m - M = 5~{\rm log}~z + 43.891
\end{equation}

\noindent If, on the other hand, we adopt the exact Mattig equation
taking into account the intrinsic geometry (i.e.\ the $q_o$ value),
then equation (5) is in error by the magnitude correction of

\begin{equation}
             \Delta (m - M) =  1.086~(1-q_o)~z
\end{equation}

\noindent (Mattig 1958; Sandage 1961, 1995). This correction will be
zero for $q_o = 1$ and approximately $0.5z$ mag for $q_o = 1/2$. The
redshifts of the Postman-Lauer clusters are all smaller than $z =
0.05$, averaging $\langle z \rangle = 0.037$; therefore, this implies
an average correction of $\sim 0.02$ mag to $M$ for $q_o = 1/2$ if
equation (5) had been used.  To be consistent with our calculation of
the linear radius, we use the exact Mattig equation with $q_o = 1/2$
for $(m - M)$ instead of equation (5). Note also that the naive choice
of the distance {\it now}, i.e.\ when the light is received, as $D =
cz/H_o$ was the choice of Hubble in his discussions of the 1930's. It
is, in fact, exact for $q_o = 1$, and it is also exact for all models
in the limit of zero redshift, seen directly from the Mattig
equations.  The Mattig (1958) equation that replaces equation (5),
using $q_o = 1/2$ with $H_o = 50$, is,

\begin{equation}
      m - M = 5~{\rm log}~[2\{1 + z - {(1 + z)}^{0.5}\}] + 43.891
\end{equation}

\noindent (Sandage 1961, 1995 equ.\ 5.14). For a redshift of $z =
0.05$, the difference between equation (7) and (5) is, in fact, 0.026
mag as expected from equation (6).
     
We have used equation (7) to calculate the $m - M$ values listed in
the first column of Table 1. The absolute magnitudes at various $\eta$
values are listed for each cluster in the body of the table using the
adopted apparent magnitudes (not listed) at each $\eta$ radius and
this calculated $m - M$ modulus. The observed apparent magnitudes
(corrected for Galactic absorption and $K$ term) can, therefore, be
recovered by applying the $m - M$ value in reverse using the listed
absolute magnitude values.
     
The fourth entry in the first column, $f$, is the log of the number
which, when added to log of the observed angular radius in arcseconds,
gives the log of the {\it linear} radius in parsecs. To calculate the
true linear radius from the observed angular radius, we must use {\it
the distance at the time that light left the galaxy}, not the distance
now when the light is received. Hence, the naive distance of $cz/H_o$
must be divided by $(1 + z)$ even in the case where we neglect the
geometry. However, we must choose the geometry in order to apply the
exact Mattig equation for ${\rm R_o} r$ (the distance now) to obtain
${\rm R_1} r$ when light left. The relation between these two proper
distances is, of course, given by the famous Lemaitre (1927, 1931)
equation

\begin{equation}
                 {\rm R_o} r/{\rm R_1} r = 1 + z
\end{equation}

\noindent derived in all the standard text books.
     
For $H_o = 50$, it follows from the exact Mattig equation (Sandage
1995, eq. 5.13a) that the linear radius\footnote{In an unfortunate
problem of nomenclature, the proper distances of ${\rm R_1}$ and ${\rm
R_o}$ in equation (8), for instance, should not be confused with our
use of the notation of R also as the linear radius of the galaxies in
equations (9) and (11), in Table 1, and in Figures 1--4, or $R$ for
the photometric bandpass. Equation (8) is the only place where we use
the notation ``R$r$'' as the distance. The ``R'' in all other places
in the text refers either to linear radius in pc, or $R$ as the
bandpass magnitude. There should be no confusion in this matter as the
context of each the meanings should be clear from the text.}, ${\rm
R(pc)}$, is given by

\begin{equation} 
  {\rm log~R(in~pc)} = f + {\rm log}~r ({\rm observed~in~arcsec})
\end{equation}

\noindent where 

\begin{equation}
   f =  {\rm log}~[(1 + z)^{0.5} - 1] - {\rm log}~(1 + z)^{1.5} + 4.765
\end{equation}

\noindent for $H_o = 50$ and $q_o = 1/2$.

The $f$ values listed as the fourth entry in column 1 of Table 1 are
calculated from equation (10). The listed log R values (the linear
radius in pc) for each $\eta$ value is given as the middle entry for
every $\eta$ value calculated from equations (9) and (10).

An example is cluster A76 in Table 1 with a redshift of $z =
0.0378$. The angular radius from the observed data (not shown) is $r =
$15\farcs{1} for $\eta = 1.3$. The listed redshift of $z = 0.0378$
gives $f = 3.013$ from equation (10), as in Table 1. Hence, log R =
3.013 + log (15.1) = 4.192, as shown in Table 1 for $\eta = 1.3$.  The
third listing in the body of the table for every $\eta$ value is the
{\it observed} average surface brightness calculated from equation
(4).  A sanity check (actually a check of the arithmetic) is available
by comparing the observed average surface brightness for each cluster
at given $\eta$ radii with the average surface brightness calculated
from the linear radius and the absolute magnitude, all in Table 1.  A
derivation of the relation between $\langle SB \rangle$, $M$, and R is
as follows. In the limit of zero redshift, using equations (7), (9),
(10), and the definition of $\langle SB \rangle = 2.5~{\rm log}~L/{\rm
R}^2$ + constant at zero redshift, it follows that the mean surface
brightness defined by equation (4) reduces to

\begin{equation}
              \langle SB \rangle = M + 5~{\rm log~R}({\rm pc}) + 22.815
\end{equation}

\noindent at zero redshift. 
     
However, applying equation (11) to the data in Table 1 does not
reproduce the listed observed surface brightnesses in line 3 for each
cluster, as calculated from equation (4). For example, for cluster A76
with log R $= 4.192$, $M = -23.37$, and the observed $\langle SB
\rangle =$ 20.57 mag per arcsec$^2$ for $\eta = 1.3$, equation (11)
predicts ${\langle SB \rangle}_{z=0} = 20.41$ mag for the $R$ band
surface brightness whereas the listed observed value is 20.57 obtained
from equation (4). This listed observed value is fainter by 0.16 mag
than that given by equation (11).

{\it However, cluster A76 is not at zero redshift}, whereas equation
(11) is valid only in the limit of $z = 0$. The reason is that
equations (7) and (10) already have built into them the Tolman signal
of $(1 + z)^4$. Therefore, equation (11) will differ from the observed
$\langle SB \rangle$ for all galaxies at redshifts larger than zero by
the Tolman signal itself of $(1 + z)^4$. Said differently, if the
Tolman signal is present, then equation (11) for any non-zero redshift
is not correct. The correct equation is equation (11) with the
$2.5~{\rm log}~(1 + z)^4$ factor added at the right\footnote{The
verification that the factor $2.5~{\rm log}~(1 + z)^4$ is needed in
equation (11) is made by forming the predicted $\langle SB \rangle$ by
using the exact Mattig equations (7), (9), and (10) for $(m - M)$ and
R as a function of the observed magnitude, the angular radius, and the
$f$ factor. It is, of course, evident that equation (11) is the
asymptotic limit of $\langle SB \rangle$ as z approaches zero.}.

The proof in the case of cluster A76 is that at its redshift of $z =
0.0378$, the predicted Tolman factor is $2.5~{\rm log}~(1.0378)^4 =
0.16$ mag, which is the difference between equation (11) and the
listed observed $\langle SB \rangle$.

This disagreement between the surface brightnesses in Table 1 entries
and those from equation (11) by the factor of $2.5~{\rm log}~(1 +
z)^4$ mag, of course, proves nothing about the validity of the Tolman
factor that is built into the standard model. It merely states that
Table 1, which uses galaxies at non-zero redshifts, does not provide
the fiducial relations for zero-redshift galaxies needed to compare
with the data for high-redshift galaxies in Papers III and IV.

In Paper IV we will correct the observed and listed $\langle SB
\rangle$ values from Table 1 by the mean redshift factor of $2.5~{\rm
log}~(1 + \langle z \rangle)^4$ for the PL cluster data to produce a
fiducial $\langle SB \rangle$ versus log R correlation that is exact
for zero redshift (see equations 12 and 13 below). The mean redshift
of the PL sample given in Table 1 is $\langle z \rangle = 0.037$ with
an rms of 0.011. The distribution of redshifts is highly peaked
between $z = 0.03$ and $z = 0.05$ because the PL sample is nearly
complete in the distance-limited sense.  This means that the
distribution of redshifts is non-linear, going as $N(z)dz \sim
z^{2}dz$. Hence, use of the mean redshift to calculate a mean
correction to zero redshift for the whole sample is an excellent
approximation. We, therefore, adopt a correction to $\langle SB
\rangle$ of 0.16 mag, making the observed $\langle SB \rangle$ values
brighter (see Tables 2 and 4).

\section{The Three Diagnostic Diagrams}

In Figure 2, we show the relation between the mean surface brightness
(corrected for Galactic absorption and $K$ term by the precepts in \S
3.3) as calculated from equation (4) and the linear radii at four
$\eta$ values for 118 of the first-ranked cluster galaxies in the PL
list. These data are taken from Table 1. The photometric data are on
the Cape/Cousins $R$ system as realized by Landolt (1983, 1992). The
linear radii are calculated from the angular radius and the galaxy
redshift by equations (9) and (10) from the Mattig equation with $q_o
= 1/2$ and $H_o = 50$; however, they are reduced to the distance when
light left the galaxy by the Lemaitre factor of $(1 + z)$. These radii
will, of course, be different if calculated using a tired light
assumption.  In that case, no expansion exists. As a consequence, no
factor of $(1 + z)$ to reduce the present distance is needed because
the distance is the same when light left as when light is
received. This case will to be treated fully in Paper IV.

We have calculated the best-fit linear least squares lines to the data
for each $\eta$ value.  The best-fit lines are shown in Figure 2, and
the parameters and their uncertainties, as determined from an
unweighted fitting to the data, are listed in Table 2.  The least
squares line for the $\eta = 2.0$ case is

\begin{equation}                 
	{\langle SB \rangle} = 2.97~{\rm log~R} + 8.69. 
\end{equation}

\noindent As discussed in \S 3.4 and listed in Table 2, the $\langle
SB \rangle$ zeropoint in equation (12) when corrected to zero redshift
is brighter by 0.16 mag, giving the relation

\begin{equation}
          {\langle SB \rangle} = 2.97~{\rm log~R} + 8.53
\end{equation}

\noindent to be used in Paper IV. 

We also list in Table 2 the adopted linear equation for $\eta = 1.0$
which we shall use in Paper IV. The data of Postman \& Lauer (1995)
are not sufficient to define the equation at this small $\eta$ value.
We have instead analyzed the extensive data for $\eta = 1.0$ from SP
III (see their Table 1).  The SP III data were taken in the $V$ band;
we have, therefore, converted to the $R$ band used in this paper by
determining the color offset from a comparison between the SP III and
the PL data at the $\eta$ values of 1.3, 1.5, 1.7, and 2.0 where the
overlap of the two data sets are adequate. Based on this comparison,
we have determined an offset of $V-R = 0.71 \pm 0.02$, which, when the
SP III data is corrected for Galactic absorption and $K$ term, is
consistent with the intrinsic colors of elliptical and S0 galaxies at
these low redshifts (Poulain \& Nieto 1994; Fukugita, Shimasaku \&
Ichikawa 1995). We note that the slopes of the correlations derived in
SP III are the same to within $\lesssim 5\%$ with those derived in
PL. This result attests to the robustness of the observational data
used in both the SP series and the PL data, each with independent
photometry.

In the final column of Table 2, we list the range of radii over which
the slope and the zero point values for the best-fit linear equations
are valid. This range is determined by the angular radii embraced by
the Postman \& Lauer (1995) data as listed in their Table 3. Figure 2
shows that the PL data do not extend to radii smaller than log R $=
4.0$ for any value of $\eta$. This lack of data at log R $< 4.0$
presents a serious problem in interpreting the high-redshift data
presented in Papers III and IV because we will need the zero-redshift
correlations for radii as small as log R $= 3.3$. In addition, SP III
(their Figures 1-6) strongly suggest that the correlation between
$\langle SB \rangle$ and log R deviates from non-linearity at smaller
radii of log R $\lesssim 4.4$ and, correspondingly, brighter surface
brightnesses.  Both of these considerations require that we provide
corrections to the linear equations in Table 2 for radii smaller than
range of radii listed there.

We have determined the corrections due to non-linearity of the slope
at log R $\lesssim 4.4$ by using the full dataset of SP III listed in
their Table 1.  The SP III data extend to metric radii of log R
$\approx 3.0$, smaller than the radii in which we are interested. The
non-linearity corrections are calculated for each $\eta$ value of 1.3,
1.5, 1.7, and 2.0. For the smallest $\eta$ value of 1.0, the data
extend only to log R $\sim 4.4$; therefore, we have simply fitted a
linear function to these data as it provides a good fit over the
observed range of radii, $3.0 \lesssim {\rm log~R} \lesssim 4.4$.

To calculate the corrections to non-linearity for the other four
$\eta$ values, we have adopted the best-fit linear relations as
calculated from the PL data (Table 2); however, we have corrected the
zeropoints to the $V$ band using the $V-R$ value given above. We then
calculate the average deviation of the data points from the best-fit
linear relation as a function of radius. We do this by averaging the
difference between the actual $\langle SB \rangle$ values and the
expected $\langle SB \rangle$ values as calculated from the linear
relation in radius bins of width $\Delta$(log R) $= 0.1$.  The
corrections to the mean surface brightness due to non-linearity are
similar for each $\eta$ value.  We have, therefore, calculated an
overall correction which is applicable to all four $\eta$ values. We
define the uncertainties on these corrections as the rms error in the
mean.  The resulting $\langle SB \rangle$ corrections as a function of
log R (in parsecs), which we shall use extensively in Paper IV, are
listed in Table 3.

Figure 3 shows the correlation of $\langle SB \rangle$ and absolute
$R$ magnitude, $M_R$, for the four values of $\eta$ = 1.3, 1.5, 1.7
and 2.0. The correlation of $\langle SB \rangle$ with absolute
magnitude, with intrinsically brighter galaxies having fainter surface
brightnesses, was suspected already in the 1960's from observations of
binary galaxies of different apparent magnitudes (Burbidge 1962;
Burbidge, Burbidge \& Crampin 1964). The linear least squares lines
are shown in each panel whose slopes and zero points are listed in
Table 4. The relation between $\langle SB \rangle$ and $M_R$ is not
linear over the full magnitude range of elliptical and S0 galaxies
(see Figure 4 of SP III); however, over the magnitude range of the PL
data, it is a reasonable approximation. We emphasize again that the
$\langle SB \rangle$ values used in Figure 3 have not been corrected
to zero redshift. To convert to zero redshift, the mean surface
brightness must be made brighter by 0.16 mag to obtain the local
calibration for zero redshift (see Table 4).

Figure 4 is the third diagnostic correlation that will be used in
Paper IV between linear radius and absolute magnitude.  The relations
in Figure 4 are similar to those in the Virgo, Fornax, and Coma
clusters presented in SP II (Figures 4 and 7), although the latter
data cover a wider range of linear radii and absolute magnitudes. A
similar correlation was found using photometric data of 56 cluster and
group galaxies in Figure 1 of SP III.

The observed change of slope as a function of $\eta$ which are shown
in Figures 2--4 and given in Tables 2 and 4 is real. This effect is
also seen in Figures 1--5 of SP III. It is one of the many current
proofs that early-type galaxies do not obey the same shape equation of
the intensity profile for all sizes and absolute luminosities as would
be the case if the classical Hubble (1930) equation with only two
parameters or the de Vaucouleurs (1948) $r^{1/4}$ equation with a
constant shape were valid for all galaxies.

King (1966) and Oemler (1976) were among the first to show that the
shapes of the intensity profiles vary systematically with intrinsic
size and absolute luminosity. Other proofs that either the Hubble or
the de Vaucouleurs constant-shape profiles cannot fit the data for all
galaxies come from studies of brightest cluster galaxies by Schombert
(1986) and, most extensively, by Graham et al.\ (1996).  Graham et
al.\ adopted a generalized $r^{1/n}$ family of modified de Vaucouleurs
profiles to fit the PL data used in this paper.  They showed that the
power-law parameter $1/n$ varied between 0.1 and 1.0, rather than
being fixed at the de Vaucouleurs value of 0.25.

Of particular relevance for us, Oemler (1976) showed that the ratio of
his outer to inner radii (as represented by his $\alpha$ and $\beta$
parameters) is a strong function of absolute magnitude (see his Tables
1 and 2).  As shown in Figure 8 of SP I, the variation of
$\alpha/\beta$ with other galaxy parameters, such as absolute
magnitude, results in a family of curves with varying ratios of
observed radius to effective radius for different $\alpha/\beta$
ratios at given $\eta$ values. By necessity, these systematic
variations of profile shape with absolute magnitude, and therefore
with physical radii, will give different slopes to the correlations in
Figures 2--4. Because of these variations, we must analyze the data in
Paper IV separately for each $\eta$ value in order to take these slope
changes into account. Our procedure in Paper IV will be to enter the
individual high-redshift data into the calibrating diagrams of Figures
2--4 separately for each $\eta$ value. We will then read, therefrom,
the depressed $\langle SB \rangle$ values at high redshift at given
log R and $\eta$ values. This depression is the Tolman signal,
modified by luminosity evolution. In this way, we generate separate
Tolman signals for each of the five $\eta$ values for which data
exists.
              
\section{Philosophy of the Test and a Summary of the Local Calibrations}

In order to test for the presence or absence of a Tolman $\langle SB
\rangle$ signal in Paper IV, we will compare the local correlations
set out in Figures 2, 3, and 4 (as corrected to zero redshift) with
the observed surface brightness, linear radii, and absolute magnitudes
of early-type galaxies at high redshift. To do this we must know how
to calculate the linear radii and absolute magnitudes used as
coordinates in these diagrams. Knowledge of how to make the
calculations depends on assumptions that we make about the
cosmological model. Here, we must introduce what may appear to be
circular reasoning.

The only coherent series of theoretical cosmological models that give
recipes on how to calculate linear radii and absolute magnitudes from
the observed data at high redshift are the standard models that assume
a metric with an expanding manifold.  These recipes due to Mattig
(1958) need specifications of the geometry to show how ``distances''
are to be calculated. They also need knowledge of the physics of how
light emitted by a distant galaxy is dimmed as it spreads over the
manifold with any particular intrinsic geometry.  All of these recipes
already assume that the universe actually expands and that the
manifold has the large-scale geometry of the
Friedman-Robertson-Walker-Mattig standard model. A summary from an
observer's point of view of how the theory interacts with the
observations in this model is set out elsewhere (Sandage 1995,
Chapters 1--6).

However, buried in these necessary recipes is the Tolman $(1 + z)^4$
factor itself. How then can we expect to make a test for its presence
if we already must use the factor to invent the recipes for the
determination of distance?  

The solution is that we also have an {\it observed} quantity for which
we need assume nothing about the cosmology in order to determine
it. This is the observed surface brightness, obtained from equation
(4) using only the observed angular radius and the observed apparent
magnitude. This observed quantity contains almost all of the Tolman
signal with only a slight dependence on $M$ and linear $R$. The Tolman
test that we make in Paper IV is for coherence in the following
way. In Paper IV we will use the observed angular radii and apparent
magnitude data from HST for the high-redshift clusters from Paper
III. We will also assume that the standard model with its explicit
recipes for R and M is correct. We will also use a variety of $q_o$
values to test the sensitivity to the intrinsic geometry.

Using the {\it observed, model free} mean surface brightness values,
we will put this $\langle SB \rangle$ data and the {\it calculated}
log (linear radii) data for the high-redshift clusters into Figure 2
to test for contradictions. If the standard model is correct with its
required $(1 + z)^4$ surface brightness dependence, the data must lie
fainter than the local fiducial line by this Tolman signal as modified
by luminosity evolution in the look-back time.  There will either be
coherence with this prediction or not. If not, there will be a
contradiction, with its consequence for the validity of Mattig
cosmology. This test for coherence removes the apparent circularity.

Anticipating the results from Paper IV, we report here that the {\it
observed} $\langle SB \rangle$ data interpreted with the standard
model as just described, using a theoretically calculated luminosity
evolution correction for redshifts between 0.7 and 0.9, show a nearly
perfect Tolman signal of four factors of $(1 + z)$ to within the
observed errors that propagate through the procedures at every
stage. We also show in Paper IV that the result is nearly independent
of $q_o$.  A tired light model is definitively excluded based on the
discrepancy between the $\langle SB \rangle$ observations at a given
linear radius in Figure 2 and the prediction of only one factor of $(1
+ z)$. No reasonable evolution correction in the look-back time can
overcome this discrepancy.

The crucial diagram for the Tolman test is Figure 2. Its agreement
with the similar correlations between $\langle SB \rangle$ and log $R$
found in the literature from a variety of sources, such as Djorgovski
\& Davis (1987) for giant cluster galaxies, SP II for galaxies in the
Virgo, Fornax, and Coma clusters, and SP III (Figure 6 and equation 1
there) that used the many sources cited in \S 1, show that the local
calibration, to which the high-redshift galaxies are to be compared,
is very well determined as we showed earlier in \S 4.

\acknowledgments 

LML was supported by NASA through Hubble Fellowship grant
HF-01095.01-97A from the Space Telescope Science Institute, which is
operated by the Association of Universities for Research in Astronomy,
Inc., under NASA contract NAS 5-26555. AS acknowledges support for
publication from NASA grants GO-5427.01-93A and GO-06459.01-95A for
work that is related to data taken with the {\it Hubble Space
Telescope}.

\newpage

\figcaption[ngc3379.ps]{Showing how the growth curve, $B(r)$, and the
Petrosian $\eta$ parameter, $\eta(r)$, are related to the profile
$SB(r)$. The profile data for NGC 3379 are from de Vaucouleurs \&
Capaccioli (1979). The solid line is an Oemler (1976) modified Hubble
profile with $\alpha/\beta = 70$. The $\eta$ values have been
calculated by the integral method in the text for every listed profile
point. This diagram is taken from Sandage \& Perelmuter (1990a).}

\figcaption[rsb.ps]{Correlations for four $\eta$ radii between log R
(linear radius in parsecs) and the observed mean surface brightness in
the Cape/Cousins $R$ band over the area bounded by the given $\eta$
radii. The radii are calculated from equation (9) and the Mattig
equation (10) of the text which assumes $q_o = 1/2$ and $H_o =
50$. The absolute magnitudes are calculated from equation (7). The
parameters of the linear least squares lines are given in Table 2. The
relation for zero redshift will be 0.16 mag brighter in $\langle SB
\rangle$ than what is listed in Table 1 and plotted in this diagram if
a Tolman signal is present in the way that equation (13) differs from
equation (12).}

\figcaption[msb.ps]{Correlation of $\langle SB \rangle$ in the $R$
photometric band with absolute magnitude, $M_R$, for four $\eta$
values. The data are listed in Table 1. The parameters of the linear
least squares lines are given in Table 4. The zero-redshift relations
will each be 0.16 mag brighter in $\langle SB \rangle$ for the same
reasons as given in Figure 2.}

\figcaption[mr.ps]{Correlation of the log R (linear radii in parsecs)
with the absolute $R$ magnitude from the data listed in Table 1. Lines
of constant surface brightness are shown. The systematic deviation of
the points from the constant surface brightness lines clearly shows
the same results as in Figure 2 that the surface brightnesses of the
intrinsically brightest galaxies are fainter than that of the
intrinsically fainter galaxies; the radius of the brightest galaxies
increases more rapidly with $M$ than $0.2~M$, which is the constant
surface brightness condition shown by the lines.}

\newpage

\begin{deluxetable}{lrrrrrr}
\scriptsize
\tablewidth{0pt}
\tabletypesize{\scriptsize}
\tablenum{1}
\tablecaption{Basic Data from Postman--Lauer on Local First Ranked
Cluster Elliptical Galaxies (Cape/Cousins $R$ Band; $H_o = 50$)}
\tablehead{
\colhead{} &
\multicolumn{6}{c}{$\eta$} \\
\colhead{Cluster} &
\colhead{1.0} &
\colhead{1.3} &
\colhead{1.5} &
\colhead{1.7} &
\colhead{2.0} &
\colhead{2.5} \\
\colhead{$z$} &
\colhead{$M({\eta})$} &
\colhead{$M({\eta})$} &
\colhead{$M({\eta})$} &
\colhead{$M({\eta})$} &
\colhead{$M({\eta})$} &
\colhead{$M({\eta})$} \\
\colhead{$m-M$} &
\colhead{log R} &
\colhead{log R} &
\colhead{log R} &
\colhead{log R} &
\colhead{log R} &
\colhead{log R} \\
\colhead{$f$\tablenotemark{a}} &
\colhead {$\langle SB \rangle$} &                  
\colhead {$\langle SB \rangle$} &                  
\colhead {$\langle SB \rangle$} &                  
\colhead {$\langle SB \rangle$} &                  
\colhead {$\langle SB \rangle$} &                  
\colhead {$\langle SB \rangle$}} 
\startdata
A76   &          &	   &	     &	       &	  &		\\
.0378 &		 &  -23.37 &  -23.79 & -24.11  &	  &		\\
36.80 &		 &  4.192  &  4.510  & 4.770   &	  &		\\
3.013 &		 &  20.57  &  21.75  & 22.73   &	  &		\\
      &		 &	   &	     &	       &	  &		\\
A119  &		 &	   &	     &	       &	  &		\\
.0450 &		 &  -24.31 &  -24.39 & -24.48  & -24.54	  &		\\
37.18 &		 &  4.619  &  4.683  & 4.758   & 4.818	  &		\\
3.083 &		 &  21.79  &  22.03  & 22.31   & 22.56	  &		\\
      &		 &	   &	     &	       &	  &		\\
A147  &		 &	   &	     &	       &	  &		\\
.0439 &		 &	   &	     &	 -23.53&   -23.76 &		\\
37.12 &		 &	   &	     &	 4.324 &   4.549  &		\\
3.074 &		 &	   &	     &	 21.08 &   22.98  &		\\
      &		 &	   &	     &	       &	  &		\\
A160  &		 &   	   &	     &	       &	  &		\\
.0443 &	   -23.05&  -23.55 &  -23.74 &	       &	  &		\\
37.14 &	   4.227 &  4.523  &  4.649  &	       &	  &		\\
3.077 &	   21.08 &  22.06  &  22.50  &	       &	  &		\\
      &		 &	   &	     &	       &	  &		\\
A168  &		 &	   &	     &	       &	  &		\\
.0453 &		 &  -23.51 &  -23.78 & -24.01  & -24.06	  &		\\
37.19 &		 &  4.266  &  4.473  & 4.676   & 4.721	  &		\\
3.086 &		 &  20.82  &  21.59  & 22.37   & 22.55	  &		\\
      &		 &	   &	     &	       &	  &		\\
A189  &		 &	   &	     &	       &	  &		\\
.0335 &		 &	   &	     & (-22.81)&  -22.96  & (-23.14)	\\
36.53 &		 &	   &	     & (4.204) &  4.382   & (4.689)	\\
2.964 &		 &	   &	     & (21.16) &  21.90   & (23.26)	\\
      &		 &	   &	     &	       &	  &		\\
A193  &		 &	   &	     &	       &	  &		\\
.0491 &		 &  -24.18 &	     &	       &	  &		\\
37.37 &		 &  4.618  &	     &	       &	  &		\\
3.118 &		 &  23.93  &	     &	       &	  &		\\
      &		 &	   &	     &	       &	  &		\\
A194  &		 &	   &	     &	       &	  &		\\
.0183 &		 &  -23.43 &  -23.62 & -23.89  &	  &		\\
35.21 &		 &  4.294  &  4.454  & 4.669   &	  &		\\
2.712 &		 &  20.93  &  21.54  & 22.35   &	  &		\\
      &		 &	   &	     &	       &	  &		\\
      &		 &	   &	     &	       &	  &		\\
A195  &		 &	   &  	     &	       &	  &		\\
.0433 &		 &	   &   -23.31&	-23.50 &  -23.72  &		\\
37.09 &		 &	   &   4.198 &	4.373  &  4.633	  &		\\
3.068 &		 &	   &   20.67 &	21.34  &  22.44	  &		\\
      &		 &	   &	     &	       &	  &		\\
A260  &		 &	   &	     &	       &	  &		\\
.0374 &		 &	   &   -24.22&	       &	  &		\\
36.77 &		 &	   &  (4.689)&	       &	  &		\\
3.009 &		 &	   &  (22.19)&	       &	  &		\\
      &		 &	   &	     &	       &	  &		\\
A261  &		 &	   &	     &	       &	  &		\\
.0469 &		 &  -23.50 &  -23.80 & -24.03  &	  &		\\
37.27 &		 & (4.260) &  4.487  &(4.685)  &	  &		\\
3.100 &		 & (20.81) &  21.65  &(22.41)  &	  &		\\
      &		 &	   &	     &	       &	  &		\\
A262  &		 &	   &	     &	       &	  &		\\
.0171 &		 &	   &	     &	       &  (-23.68)&    		\\
35.06 &		 &	   &	     &	       &  (4.634) &		\\
2.684 &		 &	   &	     &	       &  (22.37) &		\\
      &		 &	   &	     &	       &	  &		\\
A347  &		 &	   &	     &	       &	  &		\\
.0194 &		 &  -23.23 &  -23.72 &(-23.90) &	  &		\\
35.34 &		 &  4.257  &  4.617  &(4.787)  &	  &		\\
2.737 &		 &  20.95  &  22.26  &(22.93)  &	  &		\\
      &		 &	   &	     &	       &	  &		\\
A376  &		 &	   &	     &	       &	  &		\\
.0496 &		 &  -23.91 &  -24.23 &	       &	  &		\\
37.39 &		 &  4.531  & (4.777) &	       &	  &		\\
3.122 &		 &  21.77  & (22.68) &	       &	  &		\\
      &		 &	   &	     &	       &	  &		\\
A397  &		 &	   &	     &	       &	  &		\\
.0336 &		 &  -23.32 &  -23.92 &	       &	  &		\\
36.54 &		 &  4.170  &  4.615  &	       &	  &		\\
2.965 &		 &  20.49  &  22.11  &	       &	  &		\\
      &		 &	   &	     &	       &	  &		\\
A407  &		 &	   &	     &	       &	  &		\\
.0477 &	 (-23.43)&  -23.91 &  -23.95 & -24.02  &	  &		\\
37.31 &	   4.342 &  4.607  &  4.649  & 4.693   &	  &		\\
3.107 &	  (21.30)&  22.14  &  22.31  & 22.46   &	  &		\\
      &		 &	   &	     &	       &	  &		\\
A419  &		 &	   &	     &	       &	  &		\\
.0408 &		 &	   &	     &	       &   -22.80 &   -22.97	\\
36.96 &		 &	   &	     &	       &   4.356  &   4.584	\\
3.044 &		 &	   &	     &	       &   21.96  &   22.93	\\
      &		 &	   &	     &	       &	  &		\\
A496  &		 &	   &	     &	       &	  &		\\
.0327 &		 &  -24.10 &  -24.22 & -24.42  &	  &		\\
36.48 &		 &  4.594  &  4.671  & 4.829   &	  &		\\
2.954 &		 &  21.82  &  22.09  & 22.68   &	  &		\\
      &		 &	   &	     &	       &	  &		\\
A533  &		 &	   &	     &	       &	  &		\\
.0467 &		 &	   &   -23.39& (-23.72)&  -24.05  &		\\
37.26 &		 &	   &   4.298 & (4.598) &  4.948	  &		\\
3.098 &		 &	   &   21.11 & (22.28) &  23.70	  &		\\
      &		 &	   &	     &	       &	  &		\\
A539  &		 &	   &	     &	       &	  &		\\
.0291 &		 &	   &   -23.40&	-23.59 &  -23.69  &		\\
36.22 &		 &	   &   4.311 &	4.488  &  4.594	  &		\\
2.906 &		 &	   &   21.09 &	21.78  &  22.21	  &		\\
      &		 &	   &	     &	       &	  &		\\
A548  &		 &	   &	     &	       &	  &		\\
.0393 &		 &	   &   -23.50&	-23.72 &  -23.96  &		\\
36.88 &		 &	   &   4.329 &	4.549  &  4.789	  &		\\
3.029 &		 &	   &   21.12 &	22.00  &  22.96	  &		\\
      &		 &	   &	     &	       &	  &		\\
A569  &		 &	   &	     &	       &	  &		\\
.0194 &		 &	   & (-23.35)&	-23.46 & (-23.71) &		\\
35.34 &		 &	   &   4.279 &	4.374  & (4.697)  &		\\
2.737 &		 &	   &  (20.94)&	21.31  & (22.67)  &		\\
      &		 &	   &	     &	       &	  &		\\
A576  &		 &	   &	     &	       &	  &		\\
.0388 &		 &	   &	     &	       &   -23.07 &   -23.16	\\
36.86 &		 &	   &	     &	       &   4.267  &   4.404	\\
3.024 &		 &	   &	     &	       &   21.25  &   21.84	\\
      &		 &	   &	     &	       &	  &		\\
A634  &		 &	   &	     &	       &	  &		\\
.0275 &		 &	   &   -23.24&	-23.48 &	  &		\\
36.10 &		 &	   &   4.282 &	4.482  &	  &		\\
2.882 &		 &	   &   21.10 &	21.86  &	  &		\\
      &		 &	   &	     &	       &	  &		\\
A671  &		 &	   &	     &	       &	  &		\\
.0506 &		 &  -24.53 &  -24.76 & -24.88  & -24.97	  &		\\
37.44 &		 &  4.643  &  4.830  & 4.940   &(5.080)	  &		\\
3.130 &		 &  21.72  &  22.42  & 22.85   &(23.46)	  &		\\
      &		 &	   &	     &	       &	  &		\\
A779  &		 &	   &	     &	       &	  &		\\
.0225 &		 & (-23.71)&  -24.51 &(-24.68) &	  &		\\
35.66 &		 & (4.129) &  4.691  & 4.849   &	  &		\\
2.799 &		 & (19.84) &  21.85  &(22.47)  &	  &		\\
      &		 &	   &	     &	       &	  &		\\
A912  &		 &	   &	     &	       &	  &		\\
.0446 &		 &	   &	     &	 -22.92&   -23.14 &   -23.28	\\
37.16 &		 &	   &	     &	 4.295 &   4.545  &   4.762 	\\
3.080 &		 &	   &	     &	 21.56 &   22.59  &   23.53	\\
      &	   	 &	   &	     &	       &	  &		\\
      &		 &	   &	     &	       &	  &		\\
A957  &		 &	   &	     &	       &	  &		\\
.0441 &		 &  -24.53 &  -24.78 & -24.91  & -25.10	  &		\\
37.13 &		 &  4.697  &  4.875  & 5.011   & 5.198	  &		\\
3.075 &		 &  21.95  &  22.59  & 23.14   & 23.89	  &		\\
      &		 &	   &	     &	       &	  &		\\
A999  &		 &	   &	     &	       &	  &		\\
.0316 &		 &	   &	     &	 -23.49&   -23.61 &		\\
36.40 &		 &	   &	     &	 4.516 &   4.665  &		\\
2.940 &		 &	   &	     &	 22.03 &   22.66  &		\\
      &		 &	   &	     &	       &	  &		\\
A1016 &		 &	   &	     &	       &	  &		\\
.0317 &		 &	   &	     &	 -23.08&  (-23.46)&		\\
36.41 &		 &	   &	     &	 4.361 &  (4.781) &		\\
2.941 &		 &	   &	     &	 21.67 &  (23.39) &		\\
      &		 &	   &	     &	       &	  &		\\
A1060 &		 &	   &	     &	       &	  &		\\
.0115 &	   -23.20&  -24.16 &	     &	       &	  &		\\
34.20 &	   4.302 &  4.841  &	     &	       &	  &		\\
2.516 &	   21.17 &  22.91  &	     &	       &	  &		\\
      &		 &	   &	     &	       &	  &		\\
A1100 &		 &	   &	     &	       &	  &		\\
.0463 &		 &  -23.30 &  -23.58 & -23.91  &	  &		\\
37.24 &		 &  4.230  &  4.435  & 4.735   &	  &		\\
3.095 &		 &  20.86  &  21.60  & 22.77   &	  &		\\
      &		 &	   &	     &	       &	  &		\\
A1139 &		 &	   &	     &	       &	  &		\\
.0387 &		 & (-23.13)&  -23.87 &	       &	  &		\\
36.85 &		 &  4.171  &  4.708  &	       &	  &		\\
3.023 &		 & (20.70) &  22.65  &	       &	  &		\\
      &		 &	   &	     &	       &	  &		\\
A1142 &		 &	   &	     &	       &	  &		\\
.0345 &		 &	   &   -23.76&	-23.96 &	  &		\\
36.60 &		 &	   &   4.633 &	4.833  &	  &		\\
2.976 &		 &	   &   22.37 &	23.17  &	  &		\\
      &		 &	   &	     &	       &	  &		\\
A1177 &		 &	   &	     &	       &	  &		\\
.0315 &		 &  -23.94 &  -24.35 &	       &	  &		\\
36.40 &		 &  4.575  &  4.873  &	       &	  &		\\
2.938 &		 &  21.89  &  22.97  &	       &	  &		\\
      &		 &	   &	     &	       &	  &		\\
A1185 &		 &	   &	     &	       &	  &		\\
.0329 &		 &  -23.49 &  -23.64 &	       &	  &		\\
36.49 &		 &  4.303  &  4.431  &	       &	  &		\\
2.956 &		 &  20.98  &  21.47  &	       &	  &		\\
      &		 &	   &	     &	       &	  &		\\
A1213 &		 &	   &	     &	       &	  &		\\
.0467 &		 &	   &   -23.45&	-23.60 &  -23.80  &		\\
37.26 &		 &	   &   4.285 &	4.398  &  4.606	  &		\\
3.098 &		 &	   &   20.99 &	21.40  &  22.24	  &		\\
      &		 &	   &	     &	       &	  &		\\
A1228 &		 &	   &	     &	       &	  &		\\
.0365 &		 &	   &   -23.03&	-23.20 &	  &		\\
36.72 &		 &	   &   4.161 &	4.373  &	  &		\\
2.999 &		 &	   &   20.74 &	21.63  &	  &		\\
      &		 &	   &	     &	       &	  &		\\
A1257 &		 &	   &	     &	       &	  &		\\
.0343 &		 &	   &	     &	       &   -22.52 &   -22.86	\\
36.58 &		 &	   &	     &	       &   4.193  &   4.448	\\
2.973 &		 &	   &	     &	       &   21.40  &   22.34	\\
      &		 &	   &	     &	       &	  &		\\
A1267 &		 &	   &	     &	       &	  &		\\
.0326 &		 &	   &	     &	       &   -23.11 &   -23.31	\\
36.47 &		 &	   &	     &	       &   4.253  &   4.568     \\
2.953 &		 &	   &	     &	       &   21.10  &   22.48	\\
      &		 &	   &	     &	       &	  &		\\
A1308 &		 &	   &	     &	       &	  &		\\
.0511 &		 & (-23.58)&  -24.10 & -24.33  & -24.55	  &		\\
37.46 &		 &  4.164  &  4.564  & 4.759   & 4.974	  &		\\
3.134 &		 & (20.27) &  21.75  & 22.50   & 23.35	  &		\\
      &		 &	   &	     &	       &	  &		\\
A1314 &		 &	   &	     &	       &	  &		\\
.0330 &		 & (-23.42)&  -23.91 & -24.19  & -24.44	  &		\\
36.50 &		 &  4.218  &  4.578  & 4.800   &(5.058)	  &		\\
2.958 &		 & (20.62) &  21.93  & 22.76   &(23.80)	  &		\\
      &		 &	   &	     &	       &	  &		\\
A1367 &		 &	   &	     &	       &	  &		\\
.0213 &		 & (-23.34)&  -23.65 & -23.85  & -24.01	  &		\\
35.54 &		 &  4.156  &  4.368  & 4.536   & 4.746	  &		\\
2.776 &		 & (20.34) &  21.09  & 21.73   & 22.62	  &		\\
      &		 &	   &	     &	       &	  &		\\
A1631 &		 &	   &	     &	       &	  &		\\
.0461 &		 &  -23.70 &  -24.07 &(-24.43) &	  &		\\
37.23 &		 &  4.363  &  4.633  & 5.013   &	  &		\\
3.093 &		 &  21.12  &  22.10  &(23.64)  &	  &		\\
      &		 &	   &	     &	       &	  &		\\
A1644 &		 &	   &	     &	       &	  &		\\
.0468 &	   -24.25&  -24.73 & (-24.92)&	       &	  &		\\
37.27 &	   4.539 &  4.824  &  4.974  &	       &	  &		\\
3.099 &	   21.46 &  22.41  & (22.97) &	       &	  &		\\
      &		 &	   &	     &	       &	  &		\\
A1656 &		 & 	   &	     &	       &	  &		\\
.0232 &		 &  -24.00 &  -24.48 &(-24.78) &	  &		\\
35.73 &		 &  4.262  &  4.612  & 4.887   &	  &		\\
2.812 &		 &  20.22  &  21.49  &(22.57)  &	  &		\\
      &		 &	   &	     &	       &	  &		\\
A1736 &		 &	   &	     &	       &	  &		\\
.0446 &		 &  -24.03 &  -24.45 & -24.76  &(-24.95)  &		\\
37.16 &		 &  4.230  &  4.540  & 4.830   &(5.060)	  &		\\
3.080 &		 &  20.12  &  21.25  & 22.39   &(23.35)	  &		\\
      &		 &	   &	     &	       &	  &		\\
A1836 &		 &	   &	     &	       &	  &		\\
.0363 &		 &  -23.62 &  -23.86 & -24.05  & -24.26	  &		\\
36.71 &		 &  4.247  &  4.445  & 4.589   & 4.827	  &		\\
2.997 &		 &  20.58  &  21.33  & 21.86   & 22.84	  &		\\
      &		 &	   &	     &	       &	  &		\\
A1983 &		 &	   &	     &	       &	  &		\\
.0455 &		 &	   &	     &	 -23.18&   -23.35 &   -23.47	\\
37.20 &		 &	   &	     &	 4.156 &   4.343  &   4.538	\\
3.088 &		 &	   &	     &	 20.60 &   21.37  &   22.22	\\
      &		 &	   &	     &	       &	  &		\\
A2040 &		 &	   &	     &	       &	  &		\\
.0457 &	   -23.61&  -23.89 &  -23.99 & -24.04  & -24.10	  &		\\
37.21 &	   4.540 &  4.685  &  4.750  & 4.790   & 4.860	  &		\\
3.090 &	   22.09 &  22.54  &  22.76  & 22.91   & 23.20	  &		\\
      &		 &	   &	     &	       &	  &		\\
A2052 &		 &	   &	     &	       &	  &		\\
.0352 &	   -23.85&  -24.55 & (-24.77)&	       &	  &		\\
36.64 &	   4.434 &  4.824  & (4.974) &	       &	  &		\\
2.984 &	   21.28 &  22.53  & (23.06) &	       &	  &		\\
      &		 &	   &	     &	       &	  &		\\
A2063 &		 &	   &	     &	       &	  &		\\
.0355 &	   -23.34&  -23.92 &  -24.24 & -24.41  &	  &		\\
36.66 &	   4.262 &  4.602  &  4.837  & 4.987   &	  &		\\
2.987 &	   20.94 &  22.06  &  22.91  & 23.49   &	  &		\\
      &		 &	   &	     &	       &	  &		\\
A2107 &		 &	   &	     &	       &	  &		\\
.0419 &	   -23.77&  -24.38 &  -24.68 &	       &	  &		\\
37.02 &	   4.265 &  4.585  &  4.815  &	       &	  &		\\
3.055 &	   20.54 &  21.53  &  22.38  &	       &	  &		\\
      &		 &	   &	     &	       &	  &		\\
A2147 &		 &	   &	     &	       &	  &		\\
.0353 &		 &  -23.60 &  -24.02 & -24.32  &	  &		\\
36.65 &		 &  4.435  &  4.755  & 5.005   &	  &		\\
2.985 &		 &  21.54  &  22.72  & 23.67   &	  &		\\
      &		 &	   &	     &	       &	  &		\\
A2151 &		 &	   &	     &	       &	  &		\\
.0373 &		 &  -23.85 &  -24.06 & -24.21  & -24.32	  &		\\
36.77 &		 &  4.550  &  4.708  & 4.818   & 4.923	  &		\\
3.008 &		 &  21.87  &  22.45  & 22.85   & 23.27	  &		\\
      &		 &	   &	     &	       &	  &		\\
A2152 &		 &	   &	     &	       &	  &		\\
.0456 &		 &	   &   -23.31&	-23.55 &  -23.75  &		\\
37.21 &		 &	   &   4.189 &	4.449  &  4.744	  &		\\
3.089 &		 &	   &   20.64 &	21.70  &  22.98	  &		\\
      &		 &	   &	     &	       &	  &		\\
A2162 &		 &	   & 	     &	       &	  &		\\
.0325 &		 &	   &   -23.71&	-24.06 & (-24.44) &		\\
36.47 &		 &	   &   4.381 &	4.671  & (5.101)  &		\\
2.951 &		 &	   &   21.15 &	22.25  & (24.02)  &		\\
      &		 &	   &	     &	       &	  &		\\
A2197 &		 &	   &	     &	       &	  &		\\
.0307 &		 &  -24.01 &  -24.33 & -24.63  &(-24.97)  &		\\
36.34 &		 &  4.273  &  4.513  & 4.778   &(5.228)	  &		\\
2.928 &		 &  20.30  &  21.18  & 22.20   &(24.11)	  &		\\
      &		 &	   &	     &	       &	  &		\\
A2199 &		 &	   &	     &	       &	  &		\\
.0307 &	 (-23.71)&  -24.45 &  -24.82 &	       &	  &		\\
36.34 &	   4.258 &  4.698  &  4.988  &	       &	  &		\\
2.928 &	  (20.52)&  21.98  &  23.06  &	       &	  &		\\
      &		 &	   &	     &	       &	  &		\\
A2247 &		 &	   &	     &	       &	  &		\\
.0396 &		 &	   & (-23.10)&	-23.30 &  -23.58  &		\\
36.90 &		 &	   &  (4.082)&	4.267  &  4.572	  &		\\
3.032 &		 &	   &  (20.29)&	21.02  &  22.26	  &		\\
      &		 &	   &	     &	       &	  &		\\
A2572 &		 &	   &	     &	       &	  &		\\
.0423 &		 &	   &  (-23.76& (-24.05)& (-24.26) &		\\
37.04 &		 &	   &  (4.409)& (4.644) & (4.909)  &		\\
3.059 &		 &	   &  (21.27)& (22.16) & (23.27)  &		\\
      &		 &	   &	     &	       &	  &		\\
A2589 &		 &	   &	     &	       &	  &		\\
.0422 &	   -23.47&  -24.15 & (-24.37)&	       &	  &		\\
37.04 &	   4.304 &  4.698  & (4.898) &	       &	  &		\\
3.058 &	   21.04 &  22.33  & (23.11) &	       &	  &		\\
      &		 &	   &	     &	       &	  &		\\
A2593 &		 & 	   &	     &	       &	  &		\\
.0424 &	   -23.55&  -23.92 & (-24.13)&	       &	  &		\\
37.05 &	   4.308 &  4.520  & (4.680) &	       &	  &		\\
3.060 &	   20.98 &  21.67  & (22.26) &	       &	  &		\\
      &		 &	   &	     &	       &	  &		\\
A2634 &		 &	   &	     &	       &	  &		\\
.0314 &		 & (-24.13)& (-24.51)&	       &	  &		\\
36.39 &		 & (4.517) & (4.797) &	       &	  &		\\
2.937 &		 & (21.40) & (22.42) &	       &	  &		\\
      &		 &	   &	     &	       &	  &		\\
A2657 &		 &	   &	     &	       &	  &		\\
.0411 &		 &	   & (-22.65)& (-22.83)&  -23.04  & (-23.25)	\\
36.98 &		 &	   &  (3.967)& (4.097) &  4.317   & (4.647)	\\
3.047 &		 &	   &  (20.17)& (20.64) &  21.53   & (22.97)	\\
      &		 &	   &	     &	       &	  &		\\
A2666 &		 &	   &	     &	       &	  &		\\
.0277 &		 & (-23.54)&  -23.91 & -24.04  &(-24.19)  &		\\
36.12 &		 & (4.105) &  4.370  & 4.495   & 4.670	  &		\\
2.885 &		 & (19.92) &  20.88  & 21.37   &(22.10)	  &		\\
      &		 &	   &	     &	       &	  &		\\
A2717 &		 &	   &	     &	       &	  &		\\
.0492 &	  (-24.07& (-24.53)& (-24.83)&	       &	  &		\\
37.38 &	  (4.629)& (4.869) & (5.089) &	       &	  &		\\
3.119 &	  (22.10)& (22.84) & (23.64) &	       &	  &		\\
      &		 &	   &	     &	       &	  &		\\
A2731 &		 &	   &	     &	       &	  &		\\
.0312 &		 &	   &   -23.76&	-23.93 & (-24.10) &		\\
36.38 &		 &	   &   4.515 &	4.695  & (4.895)  &		\\
2.935 &		 &	   &   21.76 &	22.49  & (23.32)  &		\\
      &		 &	   &	     &	       &	  &		\\
A2806 &		 &	   &	     &	       &	  &		\\
.0272 &		 &	   &	     & (-23.31)&  -23.51  & (-23.69)	\\
36.08 &		 &	   &	     &	 4.255 &   4.478  &   4.788	\\
2.878 &		 &	   &	     & (20.90) &  21.81   & (23.18)	\\
      &		 &	   &	     &	       &	  &		\\
A2870 &		 &	   &	     &	       &	  &		\\
.0239 &		 &  -23.62 &  -23.92 &(-24.10) &	  &		\\
35.79 &		 &  4.349  &  4.554  &(4.711)  &	  &		\\
2.824 &		 &  21.04  &  21.76  &(22.37)  &	  &		\\
      &		 &	   &	     &	       &	  &		\\
A2877 &		 &	   &	     &	       &	  &		\\
.0242 &		 &  -24.23 &  -24.65 &(-24.83) &	  &		\\
35.82 &		 &  4.309  &  4.617  &(4.769)  &	  &		\\
2.829 &		 &  20.23  &  21.35  &(21.93)  &	  &		\\
      &		 &	   &	     &	       &	  &		\\
A2881 &		 &	   &	     &	       &	  &		\\
.0446 &		 &	   & (-22.70)&	-22.90 &  -23.03  & (-23.20)	\\
37.16 &		 &	   &  (4.140)&	4.240  &  4.410   & (4.680)	\\
3.080 &		 &	   &  (21.00)&	21.30  &  22.02   & (23.20)	\\
      &		 &	   &	     &	       &	  &		\\
A2869 &		 &	   &	     &	       &	  &		\\
.0318 &		 &	   &	     & (-23.03)&  -23.25  & (-23.48)	\\
36.42 &		 &	   &	     & (4.127) &  4.397   & (4.752)	\\
2.942 &		 &	   &	     & (20.56) &  21.69   & (23.23)	\\
      &		 &	   &	     &	       &	  &		\\
A2911 &		 &	   &	     &	       &	  &		\\
.0201 &		 &	   & (-22.35)& (-22.51)&  -22.65  &  -22.78	\\
35.42 &		 &	   &  (4.052)&	4.146  &  4.297   &  4.484	\\
2.752 &		 &	   &  (20.81)& (21.12) &  21.74   &  22.54	\\
      &		 &	   &	     &	       &	  &		\\
A3144 &		 &	   &	     &	       &	  &		\\
.0446 &		 &	   & (-22.91)&	-23.21 &  -23.43  & (-23.53)	\\
37.16 &		 &	   &   4.150 &	4.372  &  4.630   & (4.780)	\\
3.080 &		 &	   &  (20.84)&	21.65  &  22.72   & (23.37)	\\
      &		 &	   &	     &	       &	  &		\\
A3193 &		 &	   &	     &	       &	  &		\\
.0339 &		 &  -23.07 &  -23.30 & -23.52  & -23.72   &(-23.89)	\\
36.56 &		 &  4.052  &  4.251  & 4.469   & 4.679    &(4.834)	\\
2.969 &		 &  20.15  &  20.91  & 21.78   & 22.63    &(23.24)	\\
      &		 &	   &	     &	       &	  &		\\
A3367 &		 &	   &	     &	       &	  &		\\
.0443 &		 &	   &	     & (-22.84)&  -23.19  &		\\
37.14 &		 &	   &	     & (4.197) &  4.577	  &		\\
3.077 &		 &	   &	     & (21.14) &  22.69	  &		\\
      &		 &	   &	     &	       &	  &		\\
A3374 &		 &	   &	     &	       &	  &		\\
.0471 &		 &	   &	     &	       &   -22.98 &   -23.08	\\
37.28 &		 &	   &	     &	       &   4.227  &   4.352	\\
3.102 &		 &	   &	     &	       &   21.17  &   21.69	\\
      &		 &	   &	     &	       &	  &		\\
A3376 &		 &	   &	     &	       &	  &		\\
.0456 &		 &  -23.65 &  -24.03 & -24.24  &(-24.43)  &		\\
37.21 &		 &  4.309  &  4.589  & 4.769   &(5.049)	  &		\\
3.089 &		 &  20.90  &  21.92  & 22.61   &(23.82)	  &		\\
      &		 &	   &	     &	       &	  &		\\
A3381 &		 &	   &	     &	       &	  &		\\
.0375 &		 &  -22.92 &  -23.17 & -23.34  & -23.52	  &		\\
36.78 &		 &  4.260  &  4.460  & 4.615   & 4.785	  &		\\
3.010 &		 &  21.35  &  22.10  & 22.71   & 23.38	  &		\\
      &		 &	   &	     &	       &	  &		\\
A3389 &		 &	   &	     &	       &	  &		\\
.0262 &		 &  -23.42 &  -23.63 & -23.82  & -24.04	  &		\\
36.00 &		 &  4.162  &  4.302  & 4.449   &(4.782)	  &		\\
2.862 &		 &  20.32  &  20.81  & 21.36   &(22.80)	  &		\\
      &		 &	   &	     &	       &	  &		\\
A3395 &		 &	   &	     &	       &	  &		\\
.0482 &	   -23.41&  -24.23 &	     &	       &	  &		\\
37.33 &	   4.271 &  4.771  &	     &	       &	  &		\\
3.111 &	   20.96 &  22.64  &	     &	       &	  &		\\
      &		 &	   &	     &	       &	  &		\\
A3526 &		 &	   &	     &	       &	  &		\\
.0107 &		 &  -24.17 &  -24.37 &(-24.50) &	  &		\\
34.04 &		 &  4.455  &  4.610  &(4.733)  &	  &		\\
2.485 &		 &  20.96  &  21.54  &(22.02)  &	  &		\\
      &		 &	   &	     &	       &	  &		\\
A3528 &	      	 &	   &	     &	       &	  &		\\
.0536 &		 &  -24.49 &  -24.95 &(-25.21) &	  &		\\
37.56 &		 &  4.635  &  4.985  &(5.218)  &	  &		\\
3.153 &		 &  21.72  &  23.01  &(23.92)  &	  &		\\
      &		 &	   &	     &	       &	  &		\\
A3530 &		 &	   &	     &	       &	  &		\\
.0533 &		 &  -24.86 & (-25.03)&	       &	  &		\\
37.55 &		 &  4.851  &  4.978  &	       &	  &		\\
3.151 &		 &  22.43  & (22.90) &	       &	  &		\\
      &		 &	   &	     &	       &	  &		\\
A3532 &		 &	   &	     &	       &	  &		\\
.0547 &		 &  -24.96 &	     &	       &	  &		\\
37.61 &		 &  5.011  &	     &	       &	  &		\\
3.161 &		 &  23.14  &	     &	       &	  &		\\
      &		 &	   &	     &	       &	  &		\\
A3537 &		 &	   &	     &	       &	  &		\\
.0161 &		 &	   &   -23.58&	-24.08 &	  &		\\
34.93 &		 &	   &   4.296 &	4.737  &	  &		\\
2.658 &		 &	   &   20.78 &	22.49  &	  &		\\
      &		 &	   &	     &	       &	  &		\\
A3542 &		 &	   &	     &	       &	  &		\\
.0339 &		 &	   &	     &	       &  (-22.90)&  (-23.15)	\\
36.56 &		 &	   &	     &	       &  (4.189) &  (4.546)	\\
2.969 &		 &	   &	     &	       &  (21.00) &  (22.54)	\\
      &		 &	   &	     &	       &	  &		\\
A3553 &		 &	   &	     &	       &	  &		\\
.0474 &		 &	   &	     &	 -22.92&   -23.13 &  (-23.27)	\\
37.29 &		 &	   &	     &	 4.304 &   4.529  &  (4.754)	\\
3.104 &		 &	   &	     &	 21.61 &   22.53  &  (23.51)	\\
      &		 &	   &	     &	       &	  &		\\
A3554 &		 &	   &	     &	       &	  &		\\
.0470 &		 &  -23.74 &  -24.05 & -24.25  &(-24.43)  &		\\
37.27 &		 &  4.506  &  4.733  & 4.909   &(5.074)	  &		\\
3.101 &		 &  21.80  &  22.62  & 23.30   &(23.94)	  &		\\
      &		 &	   &	     &	       &	  &		\\
A3556 &		 &	   &	     &	       &	  &		\\
.0476 &		 &  -23.94 &  -24.25 & -24.42  & -24.54	  &		\\
37.30 &		 &  4.252  &  4.501  & 4.644   & 4.806	  &		\\
3.106 &		 &  20.33  &  21.27  & 21.81   & 22.50	  &		\\
      &		 &	   &	     &	       &	  &		\\
A5358 &		 &	   &	     &	       &	  &		\\
.0470 &	   -24.40&  -25.04 &  -25.39 &	       &	  &		\\
37.27 &	   4.451 &  4.827  &  5.081  &	       &	  &		\\
3.101 &	   20.86 &  22.10  &  23.02  &	       &	  &		\\
      &		 &	   &	     &	       &	  &		\\
A3559 &		 &	   &	     &	       &	  &		\\
.0467 &		 & (-23.83)&  -24.44 & -24.62  & -24.74	  &		\\
37.26 &		 &  4.243  &  4.643  & 4.798   & 4.958	  &		\\
3.098 &		 & (20.40) &  21.79  & 22.38   & 23.06	  &		\\
      &		 &	   &	     &	       &	  &		\\
A3560 &		 &	   &	     &	       &	  &		\\
.0117 &		 &	   &	     &	 -22.88&   -22.95 &   -23.09	\\
34.24 &		 &	   &	     &	 4.223 &   4.313  &   4.523	\\
2.523 &		 &	   &	     &	 21.10 &   21.48  &   22.39	\\
      &		 &	   &	     &	       &	  &		\\
A3562 &		 &	   &	     &	       &	  &		\\
.0483 &		 &  -24.41 &  -24.84 &(-24.99) &(-25.17)  &		\\
37.34 &		 &  4.780  & (5.082) &(5.212)  &(5.362)	  &		\\
3.112 &		 &  22.51  & (23.59) &(24.09)  &(24.66)	  &		\\
      &		 &	   &	     &	       &	  &		\\
A3564 &		 &	   &	     &	       &	  &		\\
.0484 &		 &	   &   -23.19&	-23.34 &  -23.48  &  -23.57	\\
37.34 &		 &	   &   4.163 &	4.313  &  4.473   &  4.613	\\
3.113 &		 &	   &   20.64 &	21.24  &  21.90   &  22.51	\\
      &		 &	   &	     &	       &	  &		\\
A3565 &		 &	   &	     &	       &	  &		\\
.0121 &		 &	   &   -23.89&	-24.01 &  -24.17  &		\\
34.31 &		 &	   &   4.342 &	4.459  &  4.605	  &		\\
2.537 &		 &	   &   20.69 &	21.15  &  21.72	  &		\\
      &		 &	   &	     &	       &	  &		\\
A3566 &		 &	   &	     &	       &	  &		\\
.0477 &		 &	   &	     & (-22.92)&  -23.23  & (-23.48)	\\
37.31 &		 &	   &	     & (4.075) &  4.375   & (4.847)	\\
3.107 &		 &	   &	     & (20.47) &  21.66   & (23.77)	\\
      &		 &	   &	     &	       &	  &		\\
A3570 &		 &	   &	     &	       &	  &		\\
.0365 &		 &	   &	     & (-23.10)&  -23.25  &  -23.37	\\
36.72 &		 &	   &	     & (4.175) &  4.330   &  4.500	\\
3.000 &		 &	   &	     & (20.74) &  21.36   &  22.09	\\
      &		 &	   &	     &	       &	  &		\\
A3571 &		 &	   &	     &	       &	  &		\\
.0390 &	   -24.60&  -24.95 &  -25.11 & -25.26  & -25.46	  &		\\
36.87 &	   4.581 &  4.800  &  4.930  & 5.051   & 5.248	  &		\\
3.026 &	   21.29 &  22.03  &  22.52  & 22.98   & 23.76	  &		\\
      &		 &	   &	     &	       &	  &		\\
A3572 &		 &	   &	     &	       &	  &		\\
.0398 &		 &	   &	     &	 -23.17&   -23.46 &   -23.64	\\
36.91 &		 &	   &	     &	 4.114 &   4.419  &   4.709     \\	 
3.034 &		 &	   &	     &	 20.38 &   21.62  &   22.89	\\
      &		 &	   &	     &	       &	  &		\\
A3574 &		 &	   &	     &	       &	  &		\\
.0149 &	   -23.41&  -24.05 &  -24.36 &(-24.60) &	  &		\\
34.76 &	   4.176 &  4.578  &  4.818  & 5.036   &	  &		\\
2.626 &	   20.34 &  21.71  &  22.60  &(23.45)  &	  &		\\
      &		 &	   &	     &	       &	  &		\\
A3575 &		 &	   &	     &	       &	  &		\\
.0366 &		 &	   &	     &	       &   -22.55 &   -22.73	\\
36.73 &		 &	   &	     &	       &   4.304  &   4.575	\\
3.000 &		 &	   &	     &	       &   21.94  &   23.12	\\
      &		 &	   &	     &	       &	  &		\\
A3581 &		 &	   &	     &	       &	  &		\\
.0217 &		 &  -23.30 &  -23.54 & -23.75  & -23.98	  &		\\
35.58 &		 &  4.346  &  4.534  & 4.706   & 4.934	  &		\\
2.784 &		 &  21.33  &  22.03  & 22.68   & 23.59	  &		\\
      &		 &	   &	     &	       &	  &		\\
A3656 &		 &	   &	     &	       &	  &		\\
.0192 &		 & (-24.12)&  -24.56 & -24.75  &	  &		\\
35.32 &		 &  4.533  &  4.855  & 5.008   &	  &		\\
2.733 &		 & (21.44) &  22.61  & 23.19   &	  &		\\
      &		 &	   &	     &	       &	  &		\\
A3676 &		 &	   &	     &	       &	  &		\\
.0404 &		 &	   &   -23.47&	-23.67 &  -23.78  & (-23.86)	\\
36.94 &		 &	   &   4.240 &	4.417  &  4.555   &  4.682	\\
3.040 &		 &	   &   20.71 &	21.40  &  21.98   & (22.53)	\\
      &		 &	   &	     &	       &	  &		\\
A3677 &		 &	   &	     &	       &	  &		\\
.0461 &		 &	   & (-22.78)&	-23.00 &  -23.24  &  -23.39	\\
37.23 &		 &	   &  (4.118)&	4.343  &  4.575   &  4.803	\\
3.093 &		 &	   &  (20.82)&	21.72  &  22.64   &  23.63	\\
      &		 &	   &	     &	       &	  &		\\
A3698 &		 &	   &	     &	       &	  &		\\
.0204 &		 &	   &	     &	 -23.06&   -23.32 &  (-23.51)	\\
35.45 &		 &	   &	     &	 4.183 &   4.458  &  (4.758)	\\
2.758 &		 &	   &	     &	 20.76 &   21.87  &  (23.18)	\\
      &		 &	   &	     &	       &	  &		\\
A3716 &		 &	   &	     &	       &	  &		\\
.0446 &		 &  -23.97 &  -24.18 & -24.34  & -24.47	  &		\\
37.16 &		 &  4.532  &  4.695  & 4.825   & 4.980	  &		\\
3.080 &		 &  21.69  &  22.30  & 22.79   & 23.43	  &		\\
      &		 &	   &	     &	       &	  &		\\
A3733 &		 &	   &	     &	       &	  &		\\
.0370 &		 &  -23.43 &  -23.63 & -23.79  &(-23.93)  &		\\
36.75 &		 &  4.464  &  4.604  & 4.742   &(4.894)	  &		\\
3.004 &		 &  21.77  &  22.36  & 22.89   &(23.51)	  &		\\
      &		 &	   &	     &	       &	  &		\\
A3736 &		 &	   &	     &	       &	  &		\\
.0487 &		 &  -24.23 &  -24.51 & -24.72  &(-24.91)  &		\\
37.35 &		 &  4.462  &  4.647  & 4.832   &(5.065)	  &		\\
3.115 &		 &  21.10  &  21.74  & 22.46   &(23.43)	  &		\\
      &		 &	   &	     &	       &	  &		\\
A3742 &		 &	   &	     &	       &	  &		\\
.0160 &		 &	   & (-22.88)&	-23.08 &  -23.29  & (-23.47)	\\
34.92 &		 &	   &  (4.156)&	4.338  &  4.579   & (4.824)	\\
2.656 &		 &	   &  (20.76)&	21.49  &  22.49   & (23.53)	\\
      &		 &	   &	     &	       &	  &		\\
A3744 &		 &	   &	     &	       &	  &		\\
.0375 &		 &	   & (-23.36)&	-23.65 &  -23.97  &		\\
36.78 &		 &	   &  (4.115)&	4.345  &  4.760	  &		\\
3.010 &		 &	   &  (20.19)&	21.05  &  22.80	  &		\\
      &		 &	   &	     &	       &	  &		\\
A3747 &		 &	   &	     &	       &	  &		\\
.0305 &		 &	   & (-23.19)&	-23.40 & (-23.71) &		\\
36.33 &		 &	   &   4.185 &	4.375  & (4.710)  &		\\
2.925 &		 &	   &  (20.68)&	21.42  & (22.79)  &		\\
      &		 &	   &	     &	       &	  &		\\
A3869 &		 &	   &	     &	       &	  &		\\
.0398 &		 &	   &	     &	 -23.25&   -23.46 &   -23.64	\\
36.91 &		 &	   &	     &	 4.194 &   4.449  &   4.684	\\
3.034 &		 &	   &	     &	 20.70 &   21.77  &   22.76	\\
      &		 &	   &	     &	       &	  &		\\
A4038 &		 &	   &	     &	       &	  &		\\
.0285 &		 &	   & (-23.22)&	-23.43 &  -23.63  & (-23.71)	\\
36.18 &		 &	   &  (4.182)&	4.365  &  4.559   & (4.697)	\\
2.897 &		 &	   &  (20.63)&	21.33  &  22.10   & (22.71)	\\
      &		 &	   &	     &	       &	  &		\\
A4049 &		 &	   &	     &	       &	  &		\\
.0286 &		 &	   &   -23.71&	-23.88 &  -24.03  &  -24.12	\\
36.19 &		 &	   &   4.440 &	4.589  &  4.799   &  4.919	\\
2.899 &		 &	   &   21.43 &	22.00  &  22.90   &  23.41	\\
      &		 &	   &	     &	       &	  &		\\
A4059 &		 &	   &	     &	       &	  &		\\
.0492 &	   -24.15&  -24.70 &  -24.89 &(-25.02) &	  &		\\
37.38 &	   4.399 &  4.719  &  4.857  &(4.969)  &	  &		\\
3.119 &	   20.87 &  21.92  &  22.42  &(22.85)  &	  &		\\
\enddata

\tablecomments{The parentheses indicate that the numbers are uncertain
because an extrapolation was made for the $\eta$ radii.}
\tablenotetext{a}{log R(pc) = $f + $log $r''$}
\end{deluxetable}

\begin{deluxetable}{crrcc}
\tablewidth{0pt}
\tablenum{2}
\tablecaption{Parameters of Best-Fit Linear Least Square Lines for Log
R(pc) versus $\langle SB \rangle$ at Five $\eta$ Values in the $R$
Bandpass\tablenotemark{a}}
\tablehead{
\colhead{$\eta$} &
\colhead{Slope\tablenotemark{b}} &
\colhead{Zero Point\tablenotemark{b}} &
\colhead{Zero Point} &
\colhead{Valid Range} \\
\colhead{} &
\colhead{} &
\colhead{$\langle z \rangle = 0.037$} &
\colhead{$z = 0$} &
\colhead{of log R}}
\startdata
1.0 & $2.97 \pm 0.05$ & $8.09 \pm 0.04$ & 7.93 & $>3.0$ \\
1.3 & $3.46 \pm 0.09$ & $5.89 \pm 0.12$ & 5.73 & $>4.4$ \\
1.5 & $3.39 \pm 0.07$ & $6.29 \pm 0.08$ & 6.13 & $>4.4$ \\
1.7 & $3.11 \pm 0.06$ & $7.78 \pm 0.07$ & 7.62 & $>4.4$ \\
2.0 & $2.97 \pm 0.05$ & $8.69 \pm 0.06$ & 8.53 & $>4.4$ \\
\enddata
\tablenotetext{a}{$\langle SB \rangle = a ({\rm log~R}) + b$}
\tablenotetext{b}{The uncertainties on the parameters are determined from an unweighted fitting to the data.}
\end{deluxetable}

\begin{deluxetable}{ccc}
\tablewidth{0pt}
\tablenum{3}
\tablecaption{Non-linearity Correction to $\langle SB \rangle$ from the
Equations of Table 2}
\tablehead{
\colhead{log R (pc)} &
\colhead{$\Delta \langle SB \rangle$} &
\colhead{$\sigma_{\Delta \langle SB \rangle}$}}
\startdata
4.5 & $+0.02$ & 0.04 \\ 
4.4 & $-0.05$ & 0.04 \\
4.3 & $-0.13$ & 0.04 \\
4.2 & $-0.22$ & 0.05 \\
4.1 & $-0.30$ & 0.05 \\
4.0 & $-0.39$ & 0.05 \\
3.9 & $-0.48$ & 0.06 \\
3.8 & $-0.57$ & 0.06 \\
3.7 & $-0.67$ & 0.07 \\
3.6 & $-0.76$ & 0.20 \\
3.5 & $-0.86$ & 0.20 \\
3.4 & $-0.97$ & 0.30 \\
3.3 & $-1.07$ & 0.30 \\
\enddata
\end{deluxetable}

\begin{deluxetable}{ccrr}
\tablewidth{0pt}
\tablenum{4}
\tablecaption{Parameters of the Best-Fit Linear Least Square Lines for
$M_R$ versus $\langle SB \rangle$ at Four $\eta$ Values in the $R$
Bandpass\tablenotemark{a}}
\tablehead{
\colhead{$\eta$} &
\colhead{Slope\tablenotemark{a}} &
\colhead{Zero Point\tablenotemark{a}} &
\colhead{Zero Point} \\
\colhead{} &
\colhead{} &
\colhead{$\langle z \rangle = 0.037$} &
\colhead{$z = 0$}}
\startdata
1.3 & $-2.02 \pm 0.23$ & $-27.00 \pm 0.12$ & $-27.16$ \\ 
1.5 & $-1.40 \pm 0.07$ & $-11.68 \pm 0.07$ & $-11.84$ \\ 
1.7 & $-1.36 \pm 0.07$ & $-10.43 \pm 0.06$ & $-10.59$ \\
2.0 & $-1.30 \pm 0.07$ & $-8.27  \pm 0.07$ & $-8.43$ \\
\enddata
\tablenotetext{a}{$\langle SB \rangle = a (M_R) + b$}
\tablenotetext{b}{The uncertainties on the parameters are determined from an unweighted fitting to the data.}
\end{deluxetable}

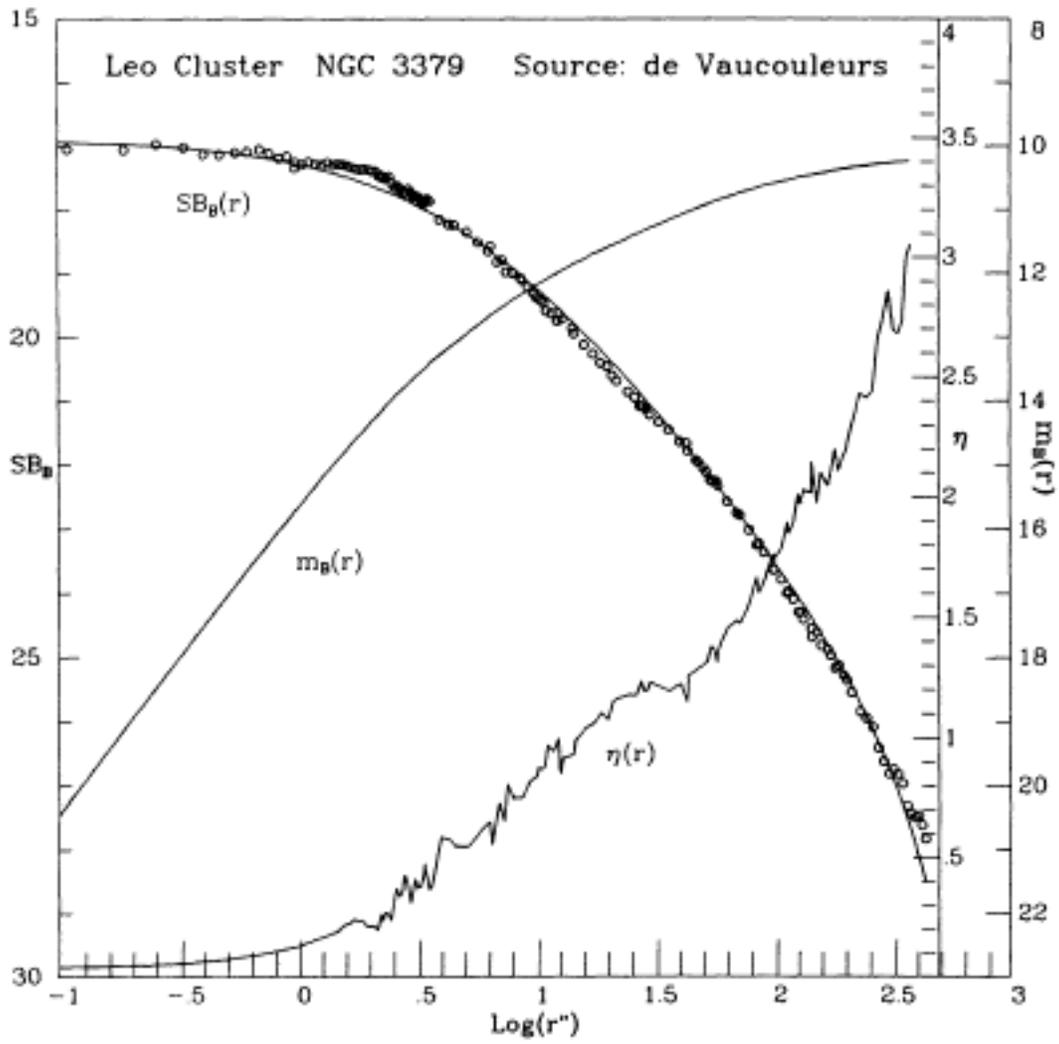
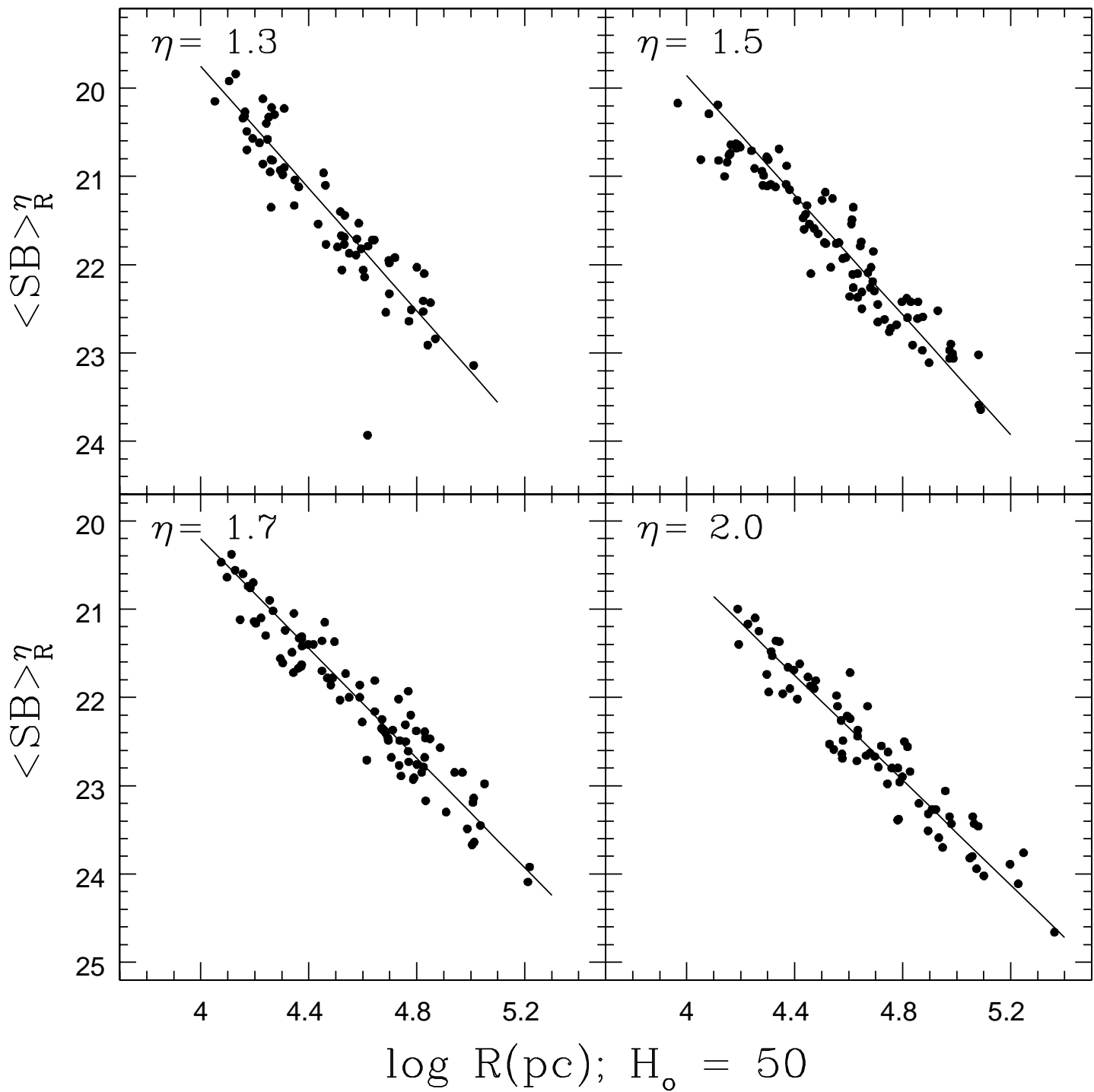
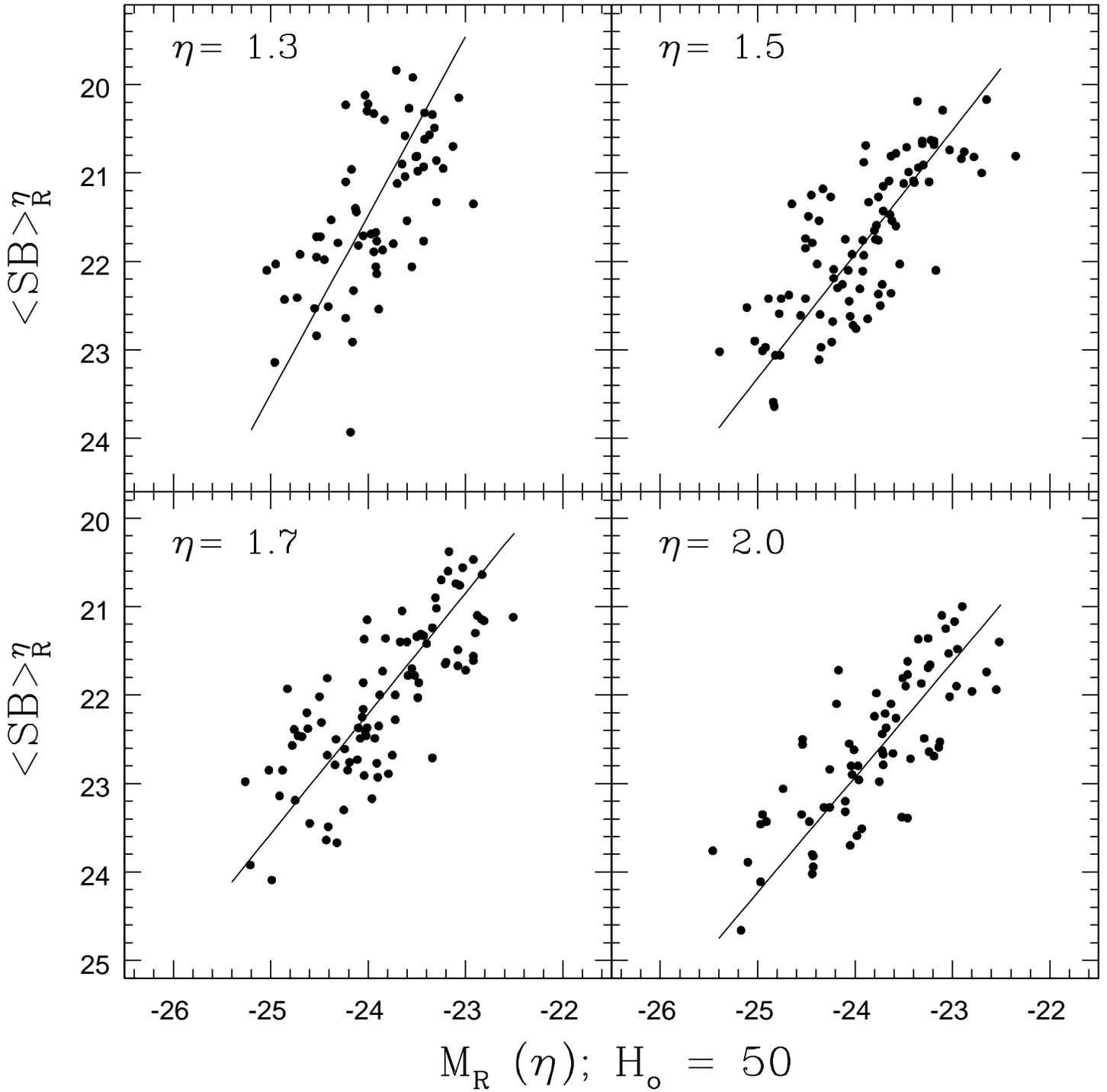
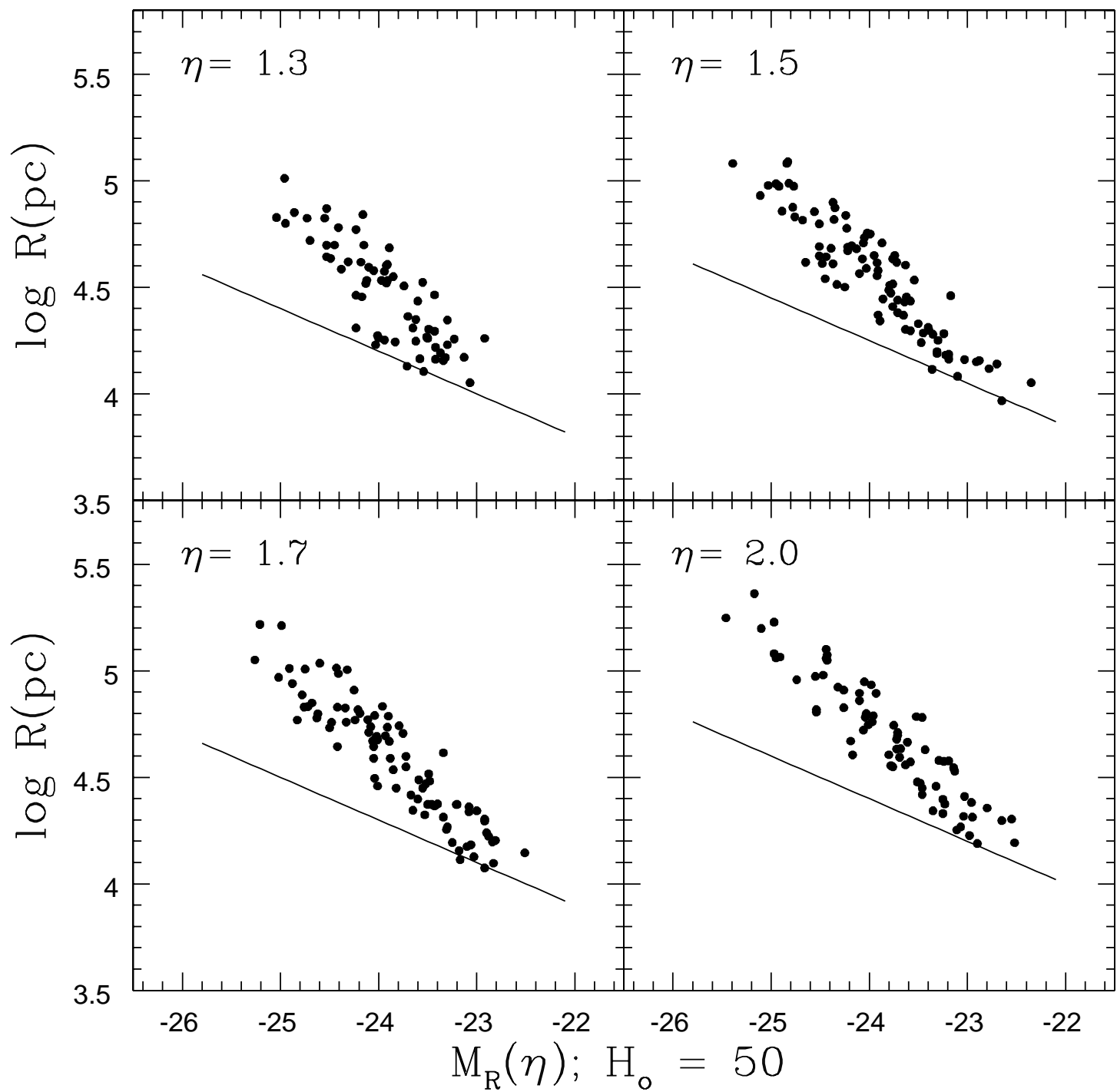
\end{document}